\documentclass[prd,a4paper,showpacs,nofootinbib,preprintnumbers,amsmath,amssymb]{revtex4}

\usepackage[dvipdfm]{graphicx}
\usepackage{amssymb}
\usepackage{amsmath}
\usepackage{parskip}
\usepackage{epsfig}
\usepackage{dcolumn}
\usepackage{rotating}
\newcommand{\be}{\begin{equation}}
\newcommand{\ee}{\end{equation}}
\newcommand{\bea}{\begin{eqnarray}}
\newcommand{\eea}{\end{eqnarray}}
\newcommand{\mx}{\mbox}
\newcommand{\mt}{\mathtt}
\newcommand{\al}{\alpha}

\newcommand{\de}{\delta}

\newcommand{\e}{\epsilon}

\newcommand{\La}{\Lambda}
\newcommand{\s}{\sigma}

\newcommand{\Om}{\Omega}

\newcommand{\rb}{\bar{r}}
\newcommand{\bM}{\bar{M}}
\newcommand{\gam}{\gamma}

\newcommand{\3}{\frac{1}{3}}

\newcommand{\stwo}{\sqrt{2}}
\newcommand{\ra}{\rightarrow}
\newcommand{\Ra}{\Rightarrow}

\newcommand{\LF}{\left(}
\newcommand{\RF}{\right)}
\newcommand{\LT}{\left[}
\newcommand{\RT}{\right]}
\newcommand{\ie}{{\it i.e.\ }}
\newcommand{\hs}{\hspace{5mm}}
\newcommand{\vs}{\vspace{5mm}\\}
\newcommand{\cO}{{\cal O}}
\newcommand{\cJ}{{\cal J}}

\begin{document}

\preprint{IGPG-07/2-1}

\title{Local Void vs Dark Energy: \\ Confrontation with WMAP and Type Ia Supernovae\\}

\author{Stephon Alexander$^a$}
\email{stephon@slac.stanford.edu}
\author{Tirthabir Biswas$^a$}
\email{tbiswas@gravity.psu.edu}
\author{Alessio Notari$^b$}
\email{notari@hep.physics.mcgill.ca}
\author{Deepak Vaid$^a, $}
\email{deepak@phys.psu.edu}

\vskip 3mm

%\begin{center}
\affiliation{
{\it $^a$ Department of Physics,\\
Institute for Gravitation and the Cosmos,\\
The Pennsylvania State University,\\
104 Davey Lab, University Park, PA,16802, U.S.A \\}
\vskip 1mm
{\it $^b$
Physics Department, McGill University, \\ 3600 University Road,
Montr\'eal, QC, H3A 2T8, Canada and \\
CERN, Theory Division, CH-1211 Geneva 23, Switzerland}
}
%\end{center}

%\maketitle

\date{\today}

\begin{abstract}
It is now a known fact that if we happen to be living in the middle of a large underdense region, then we
will observe an ``apparent acceleration'', even when any form of dark energy is absent. In this
paper, we present a ``Minimal Void'' scenario, i.e. a ``void'' with minimal underdensity contrast (of about $-0.4$)
and radius ($\sim 200-250$ Mpc/$h$) that can, not only explain the supernovae data, but also be consistent with the 3-yr WMAP
data. We also discuss consistency of our model with various other measurements such as Big Bang
Nucleosynthesis, Baryon Acoustic Oscillations and local measurements of the Hubble parameter, and also point out possible observable signatures.
\end{abstract}

%\pacs{98.80.Cq,98.80.Es,74.20.Fg}% Inflation, Cosmological Constant, and BCS Theory

\maketitle

%\keywords{Dark energy, Local Void, WMAP, SNIIa}
\section{Introduction}

One of the most baffling problems in cosmology and fundamental physics today concerns the
acceleration of the universe, as inferred from the redshifting of the type Ia supernovae.  Along
with this observation, the WMAP data and the large scale structure measurements can all  be explained
by invoking a dark fluid with negative pressure dubbed as dark energy. This has given rise to the
so-called flat $\Lambda$CDM or concordance model consisting of approximately only 4\% of visible matter (baryons), the rest being dark (approximately 3/4 dark energy and
1/4 dark matter).  However, what is this dark energy and why its  abundance should be such that it happens to be
exactly in concordance with matter density today, remains very much a mystery.

Recently,
 a few researchers have tried to take a different point of view: what if the effect of large scale structure could account for the observed luminosity to redshift behavior of type Ia supernovae ({\it i.e.} give rise to an ``apparent'' acceleration of the universe), without Dark Energy?  This question is important because an affirmative answer might obviate the need for a dark energy component/cosmological constant, which has presented a plethora of unresolved theoretical issues. Recent studies of exact solutions to the Einstein equations have, in fact, been able to reproduce the observed luminosity to redshift relation that is usually attributed to acceleration, provided that we lived in a large region (``void'') that has less matter density than the spatial average density over the cosmological Horizon~\cite{void,celerier,bmn} (see~\cite{cel-review} for a review).  One might naively conclude that this result can obviate the need for dark energy.  However, in order for the void model to be taken seriously, several key issues have to be addressed.

Firstly, the observation of small, nearly scale invariant CMB temperature fluctuations, strongly supports
the principle that our universe is homogeneous and isotropic on large scales.  In our present Universe non-linear large scale structures exist, marking  a deviation from
homogeneity; however, according to our current understanding of structure formation, ${\cal O}(1)$ non-linearities are only
expected typically at scales $\sim {\cal O}( 10 {\rm Mpc}/h )$. In this case one can again argue that the effect of these
inhomogeneities on cosmology, which is governed by the Hubble scale $\sim 3000 {\rm Mpc}/h$, would be to
too small to be significant. However, there are reasons why one could be wary of such a
conclusion.

From the theoretical point of view, the non-linear behaviour of structure formation is not a trivial issue. For instance, due to non-linear effects it is known that smaller voids
can percolate to form much larger underdense structures which occupy most of the volume of the Universe (see {\it e.g.}~\cite{varun}, according to which such a percolation has a threshold, when the density is about $50\% $ lower than the average), forming what is known as a ``cosmic web'' of superclusters and voids. Also,
we note that non-standard features on the primordial power spectrum, such as a spike at a
particular scale, or some non gaussianity may enhance the possibility of having larger structures
and voids.

Observationally speaking,  several huge nonlinear structures (notably, the  Sloan Great Wall has a length of $400 /h$ Mpc~\cite{GreatWall}) have been  revealed through surveys like SDSS and 2dF (of course, these data are only tracing the visible matter, so their interpretation in terms of total matter is subject to a bias).  It is
unclear whether the presence of these large observed objects is consistent with the present
understanding of structure formation. For example Einasto~\cite{Einasto} claims a discrepancy (by a
factor of 5) between the observed abundance of such objects (superclusters) and the values obtained
using N-body numerical simulations. Peebles has also argued that our understanding of structure formation and observed voids
are in apparent contradiction~\cite{Peebles}, and that this can be classified as a `crisis' of the $\Lambda$CDM model.
Further, there has been observational evidence for the presence of a local large
underdense ($\sim 25 \%$
 less dense) region (that extends to $\sim 200 \, {\rm Mpc}/h$) from number counts of galaxies~\cite{hole}. This represents a 4 sigma fluctuation, and would be at odds with $\La$CDM.  More recently,  there has been a claim that
 the presence of the cold spot in the CMB detected in the WMAP sky~\cite{coldCMB} is also associated with a similar Big Void in the
 large scale-structure~\cite{coldspot}.  Intriguingly, the presence of such Big Voids has also
 been advocated by~\cite{Silkinoue} in order to explain some features of the low multipole anomalies in the CMB data (in addition to the cold spot).  Finally we note that two recent papers~\cite{Schwarz,Dyer} claim a significant ($95\%$ C.L.) detection of an anisotropy in the local Hubble flow in the {\it Hubble Key Project} data~\cite{Dyer} and in the SN Ia dataset~\cite{Schwarz}. This would be a completely natural consequence of being inside a large local void~\cite{anisotropy}, since, of course, we are not expected to be exactly at the center and the void is not expected to be exactly spherical.

To summarize, the large scale structure of our universe might be richer than we thought, which can have
far reaching consequences for cosmology. However, it is fair to say that the presence of large
voids becomes more unlikely (thus requiring probably a non-conventional paradigm of structure formation), as the
size of the void and the density contrast that we consider become larger.
 This  emphasizes the need to find the
``Minimal Void (MV) Model'' {\it i.e.,} with minimal length scale and underdensity contrast that is
required to give a {\it consistent fit} to the supernovae data (the reader will easily recognize that the larger is the void, in general the better is the fit) . This is the first goal of our paper. We
find that to obtain an acceptable fit (with goodness-of-fit\footnote{The goodness-of-fit for a fit is the probability that, given a set of physical parameters, the data has a $\chi^2$ smaller or equal than the observed value.} close to 50\%) to the current supernovae data one needs ``us'' to be located
roughly centrally (with $10\%$ precision in the radial position) within an underdense region stretching upto a redshift $\sim 0.08$. If one is willing to live with a worse fit (goodness-of-fit $\sim 10\%$) then one can even go down to $z\sim 0.055$. The underdensity  that is needed is of about $\delta\sim -0.4$. Now, this is a very large region (corresponding to a radius between $160/h$ Mpc and $250/h $ Mpc). However we believe that finding a viable alternative to the presence of Dark Energy is a task which is important enough to consider such possibilities (and, as noted before, similar structures have been advocated for solving other problems in cosmology, as the low-$l$ anomalies and the cold spot in the CMB).

 As an aside we note that we obtain analytical expressions for the luminosity-redshift curve for arbitrary density profiles which are excellent approximations even when the local inhomogeneous patch extends up to $\sim 400\ Mpc/h$.

The second important issue that one has to address in the context of the MV model is whether it can
reproduce the successes of $\Lambda$CDM model for many different observations, most notably  the
WMAP third year data.   In this paper we present an analysis of the MV model subject to the WMAP 3yr data using the COSMOMC package~\cite{cosmomc} developed to perform a likelihood analysis of theoretical parameters  using a Monte Carlo Markov Chain (MCMC) method; we refine the analysis of the type Ia supernovae data and we combine them together.  We
find that using standard statistical analysis, the MV model accounts  for the WMAP and SN Ia data
while being consistent also with local measurements of the Hubble parameter.

In what follows, we will find a consistent fit to both the WMAP and SN Ia data, provided the
Hubble parameter $H_{\mt{out}}\equiv h_{\rm out}/3000 \, {\rm Mpc}^{-1}$ outside the void is very low,
$h_{\rm out}\sim 0.45$. Then the void plays the role of providing a higher value for the local
measurements of the Hubble parameter ($H_{\mt{in}}\equiv h/3000 \, {\rm Mpc}^{-1}$).  It is exactly this jump
in the Hubble parameter  that gives rise to an apparent acceleration. Additionally, in order to fit the CMB, the
primordial spectrum has to deviate from the usual nearly flat spectrum. Specifically we try the fit allowing
for running of the spectral index in the observed 7 e-folds of the CMB sky. Our best-fit has a low
spectral index with a large running.  The overall goodness-of-fit to the WMAP 3 yr data for our best-fit model is around  $26\%$ as compared  to  41\% of the $\La$CDM model.

We should clarify that although we  quote comparative statistics between MV and $\La$CDM model, it is only meant as a guide, our aim here is not to compete with the $\La$CDM model. According to the Bayesian statistical likelihood analysis of  both the supernovae and the CMB data, our best fit MV model is still disfavored by many standard deviations as compared to the concordant $\La$CDM model. Crucially however, such an inference is based on assuming a ``flat'' prior on the value of the cosmological constant. In other words it relies on the a priori assumption that  all the values of the cosmological constant are equally likely. According to the Bayesian theory, such a priori probablities are to be assigned based on theoretical prejudice. Unfortunately our understanding of the cosmological constant is rather incomplete to say the least.  As is well known, theoretical expectations suggest an enormously large value $\sim M_p^4$, and even with supersymmetry it's ``natural'' value should have been around $({\rm TeV})^4$, in obvious disagreement with our universe. Accordingly, before the discovery of our accelerated expansion, our theoretical prejudice had been to assume that the cosmological constant must in fact vanish possibly due to some symmetry or other theoretical considerations (for a recent review see for instance~\cite{nobbenhuis}). Here we take the same approach, that the ``flat prior'' assumption may actually be misleading and therefore a direct likelihood comparison between a $\La\neq 0$  model with a $\La=0$ model may not be appropriate. Rather we should focus on ``independent'' statistical quantities such as ``goodness of fit''  which can simply test whether a given theoretical model is consistent with the observational data.
% We should emphasize that in no way are we implying that the traditional approach of using a flat
%prior on the cosmological constant be abandoned. Given our lack of any fundamental understanding of the cosmological constant, the flat prior assumption seems  very reasonable and practical, but our work points out the possible pit falls of the traditional  approach.
In other words, if we had a different theoretical prejudice (for example that a non-zero cosmological constant is ``unphysical''), then we could just ask the question whether a non-homogenous matter distribution can fit the data, with an acceptable value of the goodness-of-fit. To summarize, although our model has a worse fit than $\Lambda$CDM,  in our opinion the statistics suggest that our void model in an EdS background can still be consistent with SN and CMB.

It is natural though to wonder whether one can make these fits better by including perhaps
more parameters. We consider two such possibilities in brief. In \cite{sarkar2}  the authors
obtained a slightly better fit to the WMAP data, without Dark Energy, as compared to the $\La$CDM model by including a
bump in the spectrum at some particular scale (see also~\cite{sarkar,tarun}). We discuss how this can be integrated in the MV
framework. Moreover, this idea looks particularly appealing, because the existence of a bump in the primordial spectrum could, in fact, enhance the probability of finding large voids in the present Universe (the scale of the bump happens to be roughly the scale that we need for a Minimal Void).
Next, we consider the possibility of having a slight curvature in the model. It turns
out that this also improves the fit to WMAP considerably.

Let us now come to the question of consistency between our  model and the measurements of the local
Hubble parameter. Although different observations suggest rather different values,
\be 0.55\leq
h\leq 0.8 \label{local-h} \, , \ee
 is  perhaps a fair range to consider. As we will find out, the supernovae data essentially constrains the amount of jump, $h_{\mt{out}}/h$ (or equivalently the underdensity contrast in the void), to some range.
Combining this with the WMAP analysis (which constrains $h_{\mt{out}}$) we get an allowed range for
$h$. These allowed values are definitely low, but we find that our $h$ can be as high as $0.59$
and therefore be  consistent with the local measurements of the Hubble parameter.

Finally, we briefly discuss consistency of our model with various other measurements, such as
baryon acoustic oscillations (BAO), baryon density obtained from Big Bang Nucleosynthesis (BBN),
constraints on $\s_8$ coming from weak lensing experiments, Integrated Sachs-Wolfe (ISW) effect,
etc. An important task that we leave for future is to perform an analysis of the SDSS data
including Lyman-$\al$ forests, without which one cannot really pronounce the MV model as a viable alternative to the concordant $\La$CDM model.

We now proceed as follows: in section \ref{structure}, we introduce our swiss-cheese model and
briefly discuss the non-linear structure formation captured in this model, as well as the photon propagation in this configuration. In section
\ref{MV} we explain qualitatively how the MV model can be consistent with both the supernovae and
WMAP data, as well as local measurements of the Hubble parameter.  In section \ref{supernovae}, we
perform supernovae fits for the void model. This includes finding a SN-I best-fit parameter set and
comparing it with $\chi^2$ values for the $\La$CDM model, as well as finding a combined best-fit
parameter set, which has the maximal jump (this will be needed to better fit the WMAP data) with
``acceptable'' $\chi^2$. Next in section \ref{wmap}, we perform a MCMC analysis of the WMAP data
without a cosmological constant. Again, this involves obtaining a WMAP best-fit parameter set, and
also finding a Combined best-fit model consistent with supernovae with reasonable  $\chi^2$ as
compared to the best fit ``concordance''  $\La$CDM model. In section \ref{other}, we briefly discuss
consistency of MV model with other observations such as BBN and BAO. Finally, we conclude
summarizing our findings and also pointing out unique predictions of the MV model. The appendix
contains approximate analytical solution of the trajectory, redshift and luminosity distance of a
photon in the radially inhomogeneous ``LTB'' (Lemaitre-Tolman-Biondi) metric.

%%%%%%%%%%%%%%%%%

\section{Large Scale Structure and LTB metrics}\label{structure}
As emphasized in the introduction, we are currently living in a universe with significant
inhomogeneities: non-linear structures and voids are expected on average at scales $\sim {\cal O} (10) \,
{\rm Mpc}/h$, and there is observation of structures up to much larger scales, $\sim 300 {\rm Mpc}/h$. In this
paper we will advocate that perhaps we are sitting near the centre of a ``Big Void'' spanning
a radius of $\sim 200\ {\rm Mpc}/h$ which, as we will explain, is roughly the minimal size needed to account for
the
 SN-Ia supernova data (although one can go down to values of about $150 \, {\rm Mpc}/h$ by accepting a slightly worse fit).

An accurate way to model such inhomogeneous structures/voids, which avoids any possible pit-falls
of perturbative arguments, is to use exact solutions of General Relativity that can be studied both
analytically and numerically. In particular we focus on spherically symmetric LTB
metrics~\cite{LTB} to describe ``radially'' inhomogeneous patches of any desired radius, $L$
(such metric describes the most generic spherically symmetric dust-filled spacetime; we refer to appendix
\ref{appendix} for definitions and details).
Such spherical patches can be pasted onto a homogeneous FLRW
metric consistently~\cite{reza}. It also ensures that the average density inside the spherical patch
is the same (almost exactly, see again appendix
\ref{appendix} for details) as the background density outside the patch. Thus an underdensity around the central
region is compensated by a shell-like structure near the circumference\footnote{In fact we may speculate
that the Great Sloan Wall may be indicative of such a shell-like structure, given its location, at about $250\, {\rm Mpc}/h$ away from us, and its two-dimensional shape~\cite{GreatWall}.}.

Technically, it is somewhat complicated to describe the dynamics of the LTB metric (see appendix
\ref{appendix} for details and for the choice we made for the so-called mass function),
but intuitively it is as if one had an independent scale factor
corresponding to each (comoving) radial coordinate, $r$, which is evolving  as an independent FLRW
metric with a given spatial curvature $k(r)$. {\it A priori}, $k(r)$ is an arbitrary function which also
determines the density profile. Assuming $L\ll R_H$ (the Hubble radius) one has
\be \rho(r,t)\simeq \frac{\langle\rho\rangle(t)} {1+(t/t_0)^{2/3}\e(r)} \, ,
\qquad  {\rm where} \, \, \, \langle\rho\rangle(t) \equiv \frac{M_p^2}{6 \pi t^2} \, , \qquad
{\rm and} \, \, \, \epsilon(r)\equiv 3 k(r)+ r k'(r) \, .
\label{matterdensity}  \ee

We observe that the
FLRW behaviour for the density is given by the factor $\langle\rho\rangle(t)$, while the
fluctuations are provided by the presence of $\e(r)$ in the denominator. When $\e(r)$ is close to
its maximum value we have a void, while when it is close to  its minimum, it signals an
overdensity. Note that at early times the density contrast $\de(r,t)\equiv (\rho(r,t)-\langle \rho
\rangle(t))/ \langle \rho\rangle(t)$, defined in the usual way,  grows as $t^{2/3}$, in agreement
with the prediction of cosmological perturbation theory. On the other hand at late times, when
$(t/t_0)^{2/3}\e(r)\sim \cO(1)$,  the density contrast grows rapidly (and this result is  the same
as found within the Zeldovich approximation \cite{zeldovich}). In fact, for an overdense region,
the structure ultimately collapses, as to be expected because LTB metrics cannot account for
virialization that we observe in nature. Nevertheless, for our purpose, as long as we do not reach
the collapse time, LTB metrics adequately capture the  effects of non-linear stucture formation on
photon propagation.

 Now,  we are interested in modeling a spherical void region surrounded by a compensating shell-like structure,
and this is obtained using a $k(r)$ which starts off from a maximum at $r=0$ and falls off to a constant value at $r=L$ such that
\begin{eqnarray}
 k'(0)&=&k'(L)=0 \, , \\
 k(L)&=&\frac{4\pi}{3} \Omega_k, \qquad {\rm for} \, |\Omega_k| \ll 1 \, ,
\label{junction}
\end{eqnarray}
 One can check that such an LTB metric can consistently
match to an FLRW background~\cite{reza}, with curvature abundance $\Omega_k$. In this paper we will mostly focus on a background FLRW metric which is flat. The essential reason for choosing a flat background metric is that curvature is known to be constrained to be very small in order to get a good fit of the WMAP data along with other measurements (such as measurements of the Hubble constant~\cite{Freedman}). However, in section \ref{curvature} we will present a brief discussion on how things may change in the context on the MV model if we allow for  curvature. We note in passing that  in LTB models we are considering we do not have back-reaction effects in the outside region, \ie
on the average the FLRW regions do not feel at all the presence of holes. The particular choice
of the curvature function that we employ to model the inhomogeneities and fit the supernova data is
given by
\be k(r)=k_{\mt{max}}\LT1-\LF{r\over L}\RF^4\RT^2 \label{profile} \, . \ee One can check that
Eq.(\ref{profile}) satisfies Eq.(\ref{junction}), in the case with $\Omega_k=0$. It contains two important physical parameters, $L$ and
$k_{\mt{max}}$, which correspond to the length-scale and amplitude of fluctuations respectively
\footnote{The exponent of $r/L$ has been chosen to be equal to 4, but the reader may note that any
exponent $n>1$ is good, as well. Varying $n$ one varies the width of the shell-like structure. The
larger the $n$, the flatter the void,  and narrower the structure.
However, we choose to stick only to the case $n=4$, since  it already gives us a sufficiently
flat profile for the underdense region which we found to improve the supernova fit, and anyhow the
whole analysis and discussion is not very much affected by the precise shape of the shell.}. In the
rest of the paper, this is the profile that we will focus on, although some of the analytical
results are general for any $k(r)$.

%%%%%%%%%%%%%%%%%%%%%%%%%%%%
\section{The Minimal Void Model}  \label{MV}
By now in a series of papers~\cite{void,celerier,bmn} it has been shown that a ``large'' local underdensity can
reproduce reasonably well the luminosity distance versus redshift, $D_L(z)$,  curve that one
observes, and thereby can mimic dark energy (for slightly different approaches based on inhomogeneities, see~\cite{kolb,GreciP,enqvist,golam,wiltshire}). However, the reason one is skeptical of such an
explanation is because a straightforward extrapolation of the density fluctuations observed in CMB
gives us today a scale on nonlinearity (that is, the scale in which the expected density contrast
is of ${\cal O}(1)$) of at most $\sim {\cal O} (10) /h\ {\rm Mpc}$, much too small to explain away dark energy; as we
shall see later, we need to invoke a Big Void with a radius of about $200/h$ Mpc (and with average density contrast of roughly $\langle \delta^2 \rangle \simeq 0.4$).
The probability
of having non-linear structures at larger scales  becomes progressively smaller.  Using the
conventional {\it linear} and {\it Gaussian} power spectrum for radii of about  $\sim 200 \
{\rm Mpc}/h$ the typical density contrast instead  is only of about  $0.03 \, -\, 0.05$ (for a radius of $\sim 160 \ {\rm Mpc}/h$ the typical contrast is instead about $0.06$).  However, as
argued in the introduction, one cannot take such an analysis at its face value. There are both
theoretical and observational suggestions that we might actually have larger underdensities in such voids in our universe.

Nevertheless, it is clear that the presence of large voids becomes more and more unlikely (or that
it would require a non-conventional paradigm of structure formation) as the size of the void and
the density contrast become larger and larger.

This  emphasizes the need to find the ``Minimal Void Model'' {\it i.e.,} with minimal length scale
and underdensity contrast that is required to give a consistent fit to the supernova data. This is
obtained by realizing  that the crucial evidence for acceleration comes from the fact that we
observe a mismatch between the expansion at low redshifts (between roughly $0.03\leq z \leq 0.07$)
and the expansion at higher redshifts (where supernovae are observed~\cite{RiessGold}, between roughly
$0.4\leq z \leq 1$). This situation arises because of the  current experimental status of
supernovae observations: we have very few data in the redshift range between $0.07$ and $0.4$ (the situation will
dramatically change with the coming release of the SDSS-II supernovae data~\cite{sdss2}). Thus it is not necessary to alter
the EdS $D_L(z)$ all the way up to $z\sim \cO(1)$, but a large correction to the Hubble expansion in
the local region,   $0.03\leq z \leq 0.07$, stretching up to $\sim 200 \, {\rm Mpc}/h$, may be sufficient. In
particular if we are living in a local underdensity, then we experience extra stretching as voids
become ``more void'' (that is how structure formation works) which manifests as a local Hubble expansion rate larger  than
average (outside the patch), precisely what is required to mimic
acceleration. Another way of seeing this is that all sources in the local region have a collective
radial peculiar velocity  due to the gravitational attraction of the shell-like structure, which
adds to the overall expansion.

We may also note that recently~\cite{jump} has claimed a possible detection of a jump in the
supernova Hubble diagram, exactly in the direction of having a large void. However the void radius
(about $75 \, {\rm Mpc}/h$) and the jump (about $7\%$) are smaller than what we are proposing.

Let us now  see more precisely how the MV model works. Let us start with the observation that the
$D_L(z)$ corresponding to $\La$CDM model is in good agreement with the observed supernovae~\cite{SNIA}. Thus, if
we can ensure that our MV model can approximately agree with the  $\La$CDM  $D_L(z)$ curve
both in the low and high redshift supernovae range, then we can expect to find a good fit to the
data as well. We first focus on the high redshift region, \ie outside the LTB patch.  In this
region the  $D_L(z)$ curve of the MV model basically corresponds to that of the homogeneous EdS
curve parameterized by the lower average Hubble parameter\footnote{The discrepancy
between the LTB and EdS model goes like a Rees-Sciama effect, $(L/r_H)^3$, which is $\sim \cO(10^{-5})$, according to~\cite{swiss}.
Such a correction is irrelevant for supernovae, while it could be relevant for CMB. We note, however, that~\cite{kolb,GreciP} find a larger correction in the luminosity distance. The reason for the discrepancy, however, is still unclear to us.}, $h_{\rm out}$. Further, in
this range of high redshift supernovae, the EdS curve  can run very close to the $\La$CDM model,
albeit with a different, slightly lower, Hubble parameter, $h_{\mt{out}}$ as compared to the Hubble parameter $h$ of the $\La$CDM curve. For instance, if we
compare the EdS distance ($D_E$) with the $\La$CDM distance ($D_{\La}$)~\cite{carroll}: \be {D_{E}\over
D_{\La}}\equiv {\cal R}(z)
 \, , \ee
it turns out that the ratio ${\cal R}$ does not change much in the relevant range of high-$z$
supernovae,  $0.4\leq z\leq 1$: \be {\cal R}(0.4)/ {\cal R} (1) \simeq 1.12
\, . \ee
%
%define the ``jump'' parameter ${\cal J}$ via \be {D_{E}\over D_{\La}}\equiv  \LF{h\over
%h_{\rm out}}\RF{1\over {\cal J}} \, , \ee where $h$ and $h_{\rm out}$ corresponds to the Hubble parameter
%corresponding to the  $\La$CDM model and average EdS model, then we find that ${\cal J}$ changes
%very little in the relevant range,  $0.4\leq z\leq 1$:
%\be 1.17={\cal J}(0.4)\leq {\cal J}\leq
%{\cal J}(1)=1.32 \, . \label{curlyJ}
%\ee
Moreover, the ratio ${\cal R}(z)$ itself is proportional to the ratio $h/ h_{\rm out}$. Thus, by
choosing the latter ratio appropriately, the luminosity distance of the average EdS model can be made to
approximately coincide with that of the   $\La$CDM one in the redshift range $0.4\leq z\leq 1$, and
consequently one expects that the EdS/MV model will be consistent with the high redshift
supernovae.

Next, let us look at the low redshift region. In this region, the $D_L(z)$ curve is basically
linear, the slope being given by the Hubble parameter:
\be H_0^{-1}\equiv \lim_{z\ra 0}{D_L(z)\over
z}\ = {3000\ {\rm Mpc}\over h} \, . \label{Hubble}  \ee
Thus in order for the MV model to agree with the best-fit $\La$CDM, the Hubble
parameter inside the LTB patch should coincide with the measured  local Hubble parameter. In other words, if the MV model can account for the jump, $\cJ$,  between the
locally measured Hubble parameter $h$ inside the patch, and the lower average Hubble parameter,
$h_{\rm out}$, outside the patch: \be {\cal J} \equiv {h\over h_{\rm out}} \label{jump} \, , \ee
 then we expect to have a good agreement with the supernovae data.

Thus the challenges are
\begin{itemize}
\item to quantitatively verify our above hypothesis of being able to find a good fit to the supernovae with an
appropriate ``jump''.
\item to find whether local inhomogeneities in an LTB model can account for
such jumps.
\end{itemize}
Provided  we can make this work, such an analysis will also tell us what a good range for the jump
parameter is.

As we will see later, we find a very good fit to the SN data (where we use the dataset~\cite{RiessGold}), with goodness-of-fit $\sim 50\%$, without
$\Lambda$ in the MV models. Assuming our model, a parameter estimation (with likelihood $e^{-\chi^2/2}$) gives at 95\% C.L.
the following range for the jump parameter: \be 1.17 \leq {\cal J} \leq 1.25
\label{SNJump} \, . \ee

On the other hand the fit to the WMAP data will fix the value of the global $h_{\mt{out}}$. As we
will see in section \ref{wmap}, this is the important quantity  for the photons that come from the
last scattering surface, and not for example the local $h$. The challenge for the WMAP analysis is
first to see whether one can find at all a good fit to the CMB data, without Dark Energy. It  turns out
that one can (see section \ref{wmap}),  but, crucially, a reasonable fit of the WMAP data without
$\Lambda$ requires a relatively low Hubble parameter outside the Void:
\be 0.44\leq h_{\rm out}\leq
0.47 \label{hout}
\, ,
\ee
 (at $95\%$ C.L.).

Now, these two constraints ($h_{\mt{out}}$ from CMB and the $\cJ$ from Supernovae) can be combined
together. And the third challenge now is whether we get a local value $h$ which is consistent with
local measurements of the Hubble parameter. Combining the range Eq.(\ref{hout}) with the constraints
from SN Eq.(\ref{SNJump}), we get a reasonable range of
\be 0.51 \leq h \leq 0.59 \label{hrange} \, . \ee
(see fig.~\ref{fig:wmap-sn}) and we have to compare this with the local measurements.

These local values typically vary over a wide range.  The Hubble parameter measured using supernovae~\cite{snhubble}
reads $h=0.59^{+.04}_{-.04}$, the Hubble Key Project~\cite{Freedman} measures a value of $h=0.72^{+.08}_{-.08}$ (although in~\cite{Sandage} a lower value of $h=0.62^{+0.05}_{-0.05}$ is given, with an improved treatment of Cepheids). Measurements of clusters using Sunyaev-Zeldovich distances~\cite{Reese} (which is based on data at different redshifts, up to $z\simeq 1$) gives a much
lower estimate,  $h=0.54^{+.04}_{-.03}$ (in EdS), as does measurement at high redshift ($0.3<z<0.7$) using gravitational
lensing~\cite{Lensing}: $h=0.48^{+.03}_{-.03}$ (for a more
comprehensive summary see~\cite{sarkar}). In fact, the value of $h$ estimated also seem to decrease
as one looks at sources with larger redshifts which  would be a prediction for the MV model.
However, a detailed study of this issue is well beyond the scope of our paper, but we want to
emphasize that the local value of the Hubble parameter has a large window, Eq.(\ref{local-h}) being perhaps a fair range to consider.

Clearly there is an overlap between Eq.(\ref{hrange}) and Eq.(\ref{local-h}), which is now consistent
with supernovae, WMAP and local measurements of Hubble.

This can now be used to pinpoint the underdensity contrast required in the void. As we will
analytically show in the next section (and verify numerically), the jump parameter in LTB models
does not depend on the details of the curvature (density) profile, but only on the amplitude
$k_{\mt{max}}$, or equivalently the maximal underdensity contrast at the center of the void. We find that a central
underdensity between 44\% and 58\% reproduces the relevant range
Eq.(\ref{SNJump}) of the jump parameter, and it is easy to check that this is also consistent with Eq.(\ref{local-h}) and
Eq.(\ref{hout}). Notice however that the average underdensity is always somewhat smaller than the central value, see {\it e.g.} fig.~\ref{fig:Jumpbest}.

At this point one may be concerned about the plausibility of the MV model on two different
accounts. Firstly, even if we take the observational evidence of the existence of a large
underdense region seriously~\cite{hole}, the underdensity contrast required to be consistent with
WMAP and supernovae is quite large. Secondly, the local value of the Hubble parameter is
certainly on the lower side. Both of these problems can become milder if one could obtain acceptable
fits to WMAP with slightly higher $h_{\rm out}$.  In section \ref{wmap} we  briefly  discuss how it
may be possible to evade these problems, but a  more detailed investigation of these issues is out
of the scope of this paper.

%%%%%%%%%%%%%%%%%%%%%%%%%%%%%%%%%%%
\section{Supernovae fits}\label{supernovae}
\subsection{Analytical Results}
Our aim in this section is to quantitatively fit  the supernova data (we use here the dataset from~\cite{RiessGold}) using the MV
model. In order to have better control, we decided to  perform both numerical and analytical
analysis. As explained in~\cite{swiss}, as long as $L\ll R_H$, one can find  excellent
approximations to the luminosity distance-redshift relation.  This, not only helps us physically
understand the effects of corrections coming from inhomogeneities better, but also provides us with
a non-trivial check on the numerical calculations. In the appendix we have obtained expressions for
$D_L(r)$ and $z(r)$ (which can be used to obtain $D_L(z)$ implicitly) for any general profile. We also provide the reader with a summary of all the equations necessary to reproduce the analytic approximation for $D_L(z)$ in Appendix~\ref{analyticEqs}, in a self-contained form.
Inside the LTB patch, the redshift as a function of the radial coordinate looks like \be
z\approx{2r\over 3t_0}\LT 1+2 f\LF{3k(r)/\pi}\RF\RT \, . \ee while the angular distance is simply
given by \be D_A=r\LT 1+f\LF{3k(r)/\pi}\RF\RT \, . \ee
In deriving these formulas we have used a specific choice of the radial coordinate, given in Eq.(\ref{units}) of appendix \ref{analyticEqs}, such that $r$ approximately corresponds to the proper distance today.

 The luminosity distance, in General
Relativity, is always related to the angular diameter distance~\cite{etherington} $D_A$ via \be D_L=(1+z)^2D_A \, , \ee and thus
we now have all the ingredients to obtain $D_L(z)$ inside the patch. One can easily verify that, in the
above expressions for $D_A$ and $z$, the terms outside the brackets correspond to the FLRW
results for a flat universe. $f$ is an universal function (it does not depend on the profile) defined
in the appendix, which gives us the deviation of the $D_L(z)$ curve from the FLRW result. As one can
see, our analytical results agree very well with the numerical solutions, see fig.\ref{fig:numanalytic}.

Now, one defines the Hubble parameter as the initial (z=0) slope in the $D_L-z$ plot:  using
this definition  one can obtain (see appendix \ref{thejump} for details) an exact  relation between the jump
parameter and the central density contrast: \be \cJ={h\over h_{\rm out}}=2 -(1-|\de_0|)^{1/3} \, . \ee
Surprisingly, this expression does not depend  on the specific form of the profile, and therefore
lends generality to the analysis.

%%%%%%%%%%%%%%%%%%%%%
\subsection{Numerical Analysis}
\begin{figure}
\includegraphics[width=0.48\textwidth]{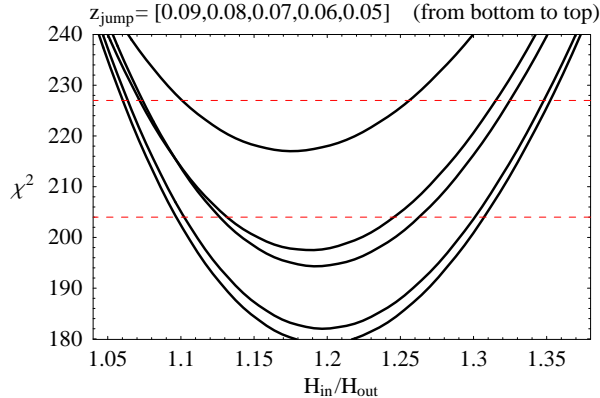}
\caption{ \label{figJump} The $\chi^2$ for Supernovae IA as a function of the jump
$h/h_{\mt{out}}=H_{\mt in}/H_{\mt out}$, for different values of the size of the inhomogeneous region (whose boundary ends at redshift
 $z_{\rm jump}$). We have used here a model with two FLRW regions (empty inside and EdS outside),
with two different Hubble parameters. From bottom to top the solid curves correspond to $z_{\rm
jump}=[0.09,0.08,0.07,0.06,0.05]$. The two dashed lines correspond to a $10\%$ and a $1\%$
goodness-of-fit. The number of d.o.f. is $181$ (we have used the Riess Gold dataset~\cite{RiessGold}). }
\end{figure}

We employ in this section a two steps strategy. First, without even using the LTB metric, we try to
fit the data with a crude approximation of the void, which consists of an empty (curvature
dominated) FLRW Hubble diagram for the inner region and then an EdS Hubble Diagram for the outer
region. Between the two regions ($z<z_{\rm jump}$ and $z>z_{\rm jump}$) there is a discontinuous jump in the
Hubble parameter $H_{\mt{in}}/H_{\mt{out}}$. In this way we get a good idea about what are the best
values for $\cJ$ and $z_{\rm jump}$. The results are shown in fig.~\ref{figJump}.

%%%%%%%%%%%%%%%%%%%%%%%%%%%%%%%%%%%%%%%%%%%%%%%%%%%%%%%%%%%%%%%%
%%%%%%%%%%%%%%%%%%%%%%%%%%%%%%%%%%%%%%%%%%%%%%%%%%%%%%%%%%%%%%%%

As one can see from the plot, the larger is the value for $z_{\rm jump}$ the better is the fit.
However, we do not gain much by taking $z_{\rm jump}$ larger than, say, $0.08$ (which corresponds to a radius of $ 250 \, {\rm Mpc}/h)$.   It is also
interesting to note  that a $z_{\rm jump}$ as low as $0.05$ (which corresponds to a radius of $150 \, {\rm Mpc}/h$) still gives a reasonable fit
(goodness-of-fit is higher than a few $\%$). Almost independent of $z_{\rm jump}$, the best value
for the jump is around $\cJ\simeq 1.2$.

As a second step, then, we try to reproduce these results with a full LTB study. For
simplicity we focus on only one value of $L$ for the LTB patch ($z_{jump}\approx 0.085$). A further
observational motivation for considering such a redshift comes from the fact that it also
approximately coincides with the redshift of the Sloan Great Wall, which  spans hundreds of Mpc
across and it could be  suggestive of being the ``compensating structure'' expected at the boundary of
the LTB patch~\cite{GreatWall}. In the  profile Eq.(\ref{profile}), we therefore fix the radius $L$, and  let $k_{\mt{max}}$ vary (which corresponds to varying the jump $\cJ$, or equivalently the central density contrast $\delta_0$).

\begin{figure}
\includegraphics[width=0.48\textwidth]{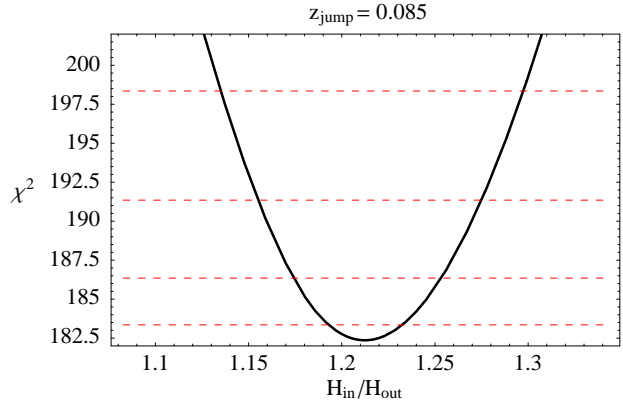}
\caption{ \label{figLTB} The $\chi^2$ for Supernovae IA as a function of the jump
$h/h_{\mt{out}}=H_{\mt in}/H_{\mt out}$, in a full specific LTB model, matched to FLRW at redshift $z_{\rm jump}=0.085 $.
The dashed lines correspond to the $1\sigma$, $2\sigma$, $3\sigma$ and $4\sigma$ where we used as a likelihood $e^{-\chi^2/2}$. The number of d.o.f. is
$181$ (we have used the Riess Gold dataset~\cite{RiessGold}). }
\end{figure}

We solve numerically for the $D_L-z$ curve for several values of $k_{\mt{max}}$ (which correspond to several
values of $\cJ$), and we compute the $\chi^2$. We show in fig.~(\ref{figLTB}) the $\chi^2$ as a
function of the jump, interpolating between the results of the numerics. This interpolating
function is then used to compute the statistics: We find that the 1$\s$ range of the  jump
corresponds to $1.214^{+.019}_{-.019}$. For the density contrast at the center this translates to
$\de_0=0.514^{+.034}_{-.036}$.

%%%%%%%%%%%%%%%%%%%%%%%%%%%%%%%%%%%%%%%%%%%%%%%%%%%%%%%%%%%%%%%%
%%%%%%%%%%%%%%%%%%%%%%%%%%%%%%%%%%%%%%%%%%%%%%%%%%%%%%%%%%%%%%%%

Let us comment briefly on the values that we get for the $\chi^2$ as compared to other models.
The EdS model has a very bad fit to  the data, since its $\chi^2$ for the same dataset is  about $284$. This has a very low goodness of fit.
On the other hand the $\Lambda$CDM model has a much lower $\chi^2$ than our model (\cite{RiessGold} quotes 150)\footnote{The open empty Universe has also a low $\chi^2$, of about 160.}, which is indeed strangely too low\footnote{We note here that all the SN fits are plagued by not knowing exactly what are the errors on the SN measurements. In fact, if one used only instrumental errors, then the data points would have a very large scatter with tiny errors, and there is no smooth curve which can give a fit to the data. Then what is done by SN collaborations is to artificially add by hand an error bar of about 0.15 magnitudes in quadrature to all data points, which is typically justified saying that this is the typical variability of the intrinsic SN luminosity. This is what makes the concordance $\Lambda$CDM $\chi^2$ so low.}.
Now, in terms of goodness-of-fit our $\chi^2$ is what one expects typically, since it is roughly equal to the number of d.o.f., and this makes our model a good fit to the data. On the other hand if one allows a new free parameter ($\Omega_{\Lambda}$) then the best fit turns out to be at a nonzero value for $\Omega_{\Lambda}$, and so the parameter value $\Lambda=0$ would be formally excluded at several $\sigma$ (assuming a likelihood that goes as $e^{-\chi^2/2}$).
This situation is similar to what we will encounter when we perform the CMB fits (see section \ref{wmap}): the MV model has a worse $\chi^2$ as compared to $\La$CDM, but the question that we want to ask  is about consistency of SN data with a MV model, and for this question the answer seems to be yes, the $\chi^2$/d.o.f. being roughly equal to 1.

We also note that we use only one dataset~\cite{RiessGold} (while there are other ones in the literature), since we would qualitatively get very similar answer and it is not our purpose here to compare dataset with others, but just to check the consistency of the model.

\begin{figure}\label{fig:Jumpbest}
\includegraphics[width=0.76\textwidth]{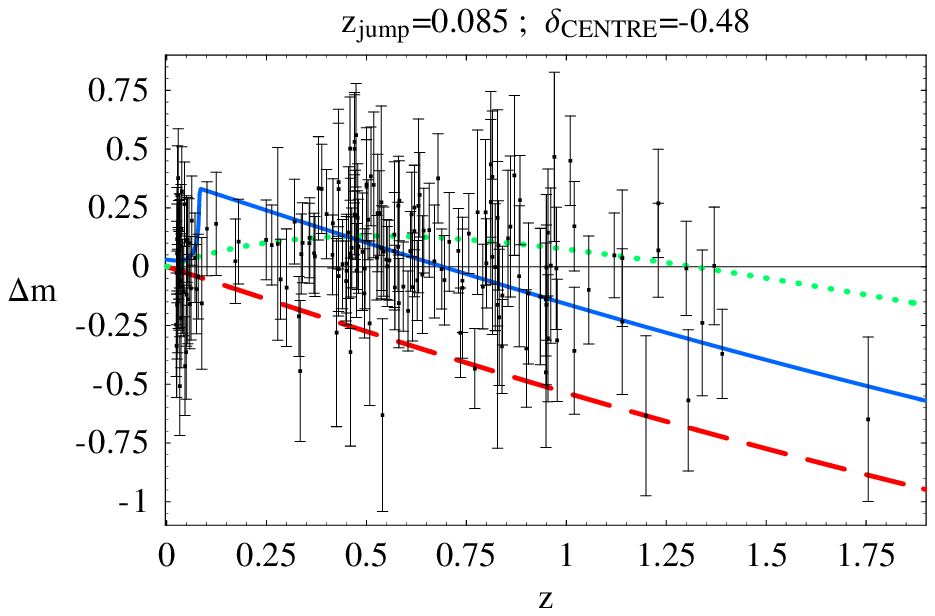}
\includegraphics[width=0.5\textwidth]{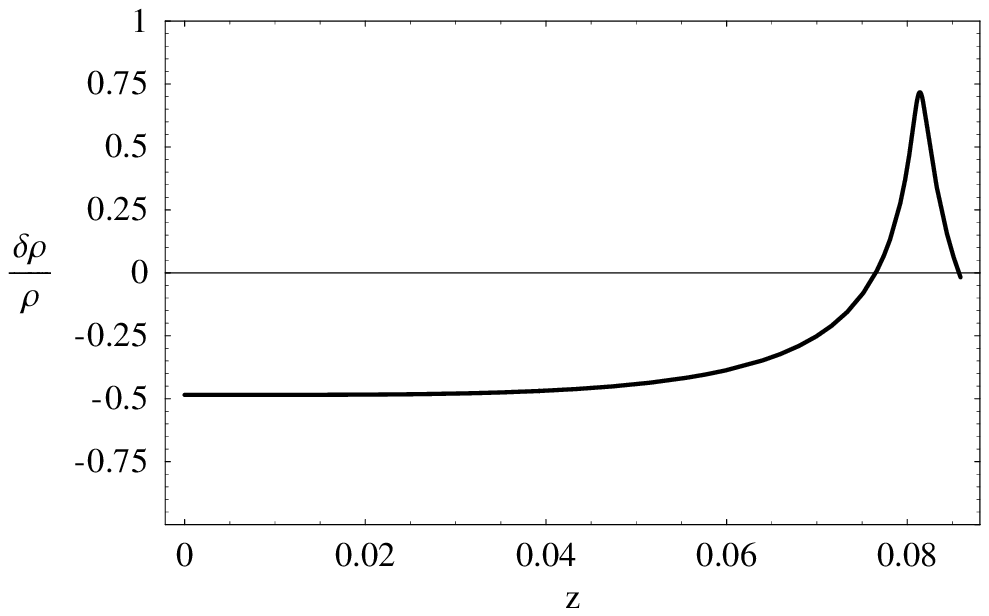}
\caption{ In the upper plot we show a fit of the Supernovae data (Riess et al.~\cite{RiessGold})
with an LTB model which has $\chi^2=186$ (the d.o.f. are 181). The inhomogeneous patch extends up to
$z\simeq 0.085$ and the underdensity in the center is $\delta_{\rm CENTRE}=-0.48$. We have shown
$\Delta m\equiv m-m_{empty}$: the magnitude ($m\equiv 5 Log_{10} D_L$)  minus the magnitude of an
empty open FLRW Universe as a function of the redshift $z$. The blue solid line is our
inhomogeneous model, the red dashed-line is an EdS model (whose Hubble constant is normalized
through the nearby supernovae), the green dotted line is the best-fit $\Lambda CDM$.  In the
lower plot we show the density contrast for the same model, as a function of $z$. The average
contrast ($\sqrt{\langle \delta^2 \rangle}$) in the inhomogeneous patch is $0.43$ ($\sqrt{\langle
\delta^2 \rangle}\simeq 0.33$ in the underdensity,  $\sqrt{\langle \delta^2 \rangle}\simeq 0.48$ in
the overdensity). }
\end{figure}

Finally we show, as an illustration,  one example of a plot of $D_L-z$ in
figure~\ref{fig:Jumpbest} together with the shape of the density profile (as a function of $z$) .

%%%%%%%%%%%%%%%%%%%%%%%%%%%%%%%%%%%%%%%%%%%%%%%%%%%%%%%%%%%%%%%%
%%%%%%%%%%%%%%%%%%%%%%%%%%%%%%%%%%%%%%%%%%%%%%%%%%%%%%%%%%%%%%%%

%%%%%%%%%%%%%%%%%%%%
\section{MCMC fit of the WMAP data}\label{wmap}

In order for the MV model to be viable at all, it is crucial for this picture to be in agreement
with observations of the CMB spectra, among other things. It is commonly assumed that the
$\Lambda$CDM model, with a non-zero cosmological constant,  is the only one which can adequately
explain the CMB spectrum. This is based on the result that once one assumes a ``flat'' prior on
$\Lambda$, it turns out that the ``most likely'' parameters,  given the WMAP data, correspond to
$\Om_{\La}\sim 0.7$. The question that we want to ask  however, is about consistency of WMAP with
EdS: can we get a reasonable fit to the CMB spectrum even after setting $\Lambda$ to zero? To put it differently, if we had a strong theoretical prejudice
against having a non-zero cosmological constant, or if there were other observations  disfavoring
it, then would  the 3-yr WMAP data independently rule out an $\Om_M=1$, EdS universe? (Here $M$
means total matter = baryons + dark matter.) In particular, what if we introduce additional
features in the primordial spectrum, rather than tampering with the composition of the
universe?  If we indeed obtain a reasonable fit using such a different ``prior'', the next
important step would be to check whether this parameter set is consistent with the supernovae fit.
This is what we plan to do in this section.

Rigorously speaking, this question seems technically challenging because one would have to compute the secondary effects, {\it i.e.} what
the spectrum of the CMB radiation would look like after passing through the local underdense
region, and maybe many other such regions\footnote{The assumption that we live in a void could
naturally lead us to consider that the universe might  contain many such voids, a bubbly
universe. In this case one would have to compute the passage of the photons through many such voids.}, it
encounters on its journey to us.   According to~\cite{swiss}, the corrections to the redshifts of photons
which pass through a void of size $L$ is a Rees-Sciama effect that goes like $(L/R_H)^3$.  A coherent addition of this effect
due to many voids could produce a correction of order $ (L/R_H)^2$. Thus for a void with a
typical radius $\sim 200/h$ Mpc that we considered in this paper, such a cumulative effect could be $\sim 10^{-2}-10^{-3}$.  This can be ignored for the study of supernovae\footnote{We mention, again, that~\cite{GreciP, kolb} find a larger correction to the luminosity distance, which could be potentially important for supernovae~\cite{kolb}. Since the reason of the discrepancy is still unclear to us, we do not discuss it here.}.
On the other hand,
if these many voids exist, they would give a sizable effect on the CMB. The number $\sim
10^{-2}-10^{-3}$ would refer to a monopole in the CMB, while the correction to higher multipoles would be smaller
(depending on how different is the number of voids along different directions in the sky). However, in this paper we ignore such secondary effects. On the qualitative
side, in fact, we expect this to be important only for small $l$ and decay fast for larger $l$, and it should
act in the same way as an Integrated Sachs-Wolfe effect~\footnote{This might explain the claimed detections of
ISW correlations~\cite{isw}, even without invoking Dark Energy.}.

The correction to the CMB redshift that comes from our local void, instead, will depend on how
symmetric the void is, and how ``centrally''  we are located. For an off-center observer, in appendix \ref{dipole} we perform a non-perturbative estimate of the dipole moment, and find that in order for it to not exceed the observed value $\sim\cO(10^{-3})$, ``we'' must be located very close to the center, approximately within 10\% of the void-radius, in concordance with the findings in~\cite{alnes-dipole}. In this case the correction to the higher multipoles are much more suppressed and not visible in CMB~\cite{alnes-dipole}. Departure from spherical symmetry, on the other hand, may have a  much more interesting effect, specially on the lowest $l$s in the CMB spectrum, and could be visible\footnote{In this context we note that similar effects in anisotropic geometric void configurations  have been used to explain the low multipole anomalies in
the CMB sky~\cite{Silkinoue}.}. However, such a study is clearly out of the scope of the present paper;  instead we will restrict  ourselves to  spherically symmetric voids and  neglect these possible  secondary effects on the CMB coming from  the  voids
embedded in the homogeneous EdS background. Thus, the question
reduces to whether the CMB spectrum can be reproduced given an EdS background.

As one would expect, we find that if one assumes as priors, no dark energy, as well as  no
additional features in the primordial spectrum (other than spectral index and amplitude), one obtains a very
poor fit to the 3-yr WMAP data (see table \ref{table:fits}). However the situation changes if we introduce a possible ``running in the spectral tilt'', $\al_s$, in the observable
$\sim 7$ e-folds of our universe in CMB (following the same definition as in~\cite{cosmomc})
\footnote{Ideally, one would like to introduce two
additional scales where significant running of the spectral tilt starts and ends. This would also
obviously improve the fit. However, to keep the analysis simple, we have assumed that these two
scales lie outside the observed spectrum in WMAP.}

 We have performed a Monte Carlo Markov Chain (MCMC) analysis of the WMAP 3 year data
using the program COSMOMC \cite{cosmomc}. Our runs  were performed with the
priors given in table~\ref{tabellapriors}.
%\vs
\begin{table} \label{tabellapriors}
\begin{tabular}{|c|c|c|c|c|c|c|c|c|}
\hline
    &\multicolumn{2}{c|}{\bf{$\Lambda$CDM}} &\multicolumn{2}{c|}{\bf{EdS}}
    &\multicolumn{2}{c|}{\bf{EdS}} &\multicolumn{2}{c|}{\bf{Curved}}
% &\multicolumn{2}{c|} {\bf{Open,$\al_s\neq0$}}&\multicolumn{2}{c|}{\bf{Closed,$ \alpha_s\neq0$}}
\\
     &\multicolumn{2}{c|}{\bf{}} &\multicolumn{2}{c|}{\bf{$ \alpha_s=0$}}
    &\multicolumn{2}{c|}{\bf{$ \alpha_s\neq 0$}} &\multicolumn{2}{c|}{\bf{$ \alpha_s, \Omega_k\neq 0 $}}
% &\multicolumn{2}{c|}{\bf{$\Omega_k=-0.05 $}}&\multicolumn{2}{c|}{\bf{$\Omega_k=0.05 $}}
\\ \hline
        &\emph{min}    &\emph{max}    &\emph{min}    &\emph{max}    &\emph{min}    &\emph{max}    &\emph{min}    &\emph{max}
% &\emph{min}    &\emph{max}  &\emph{min}    &\emph{max}
\\ \hline
$ \Omega_b h_{\mt{ out}}^2 $    &$0.005$    &$0.04$ &$0.005$    &$0.04$ &$0.005$ &$0.04$
&$0.005$    &$0.04$
% &$0.005$    &$0.04$ &$0.005$    &$0.04$
\\ \hline

$ \Omega_{m} h_{\mt{out}}^2 $ &$0.01$ &$0.3$  &$0.01$ &$0.3$  &$0.01$  &$0.3$ &$0.01$  &$0.3$
% &$0.01$  &$0.3$ &$0.01$   &$0.3$
\\ \hline

$ \Omega_{\La} $    &$0$ &$1$  &$0$    &$0$    &$0$    &$0$    &$0$ &$0$
% &$0$ &$0$ &$0$ &$0$
\\ \hline
$ n_s $ &$0.5$  &$1.5$  &$0.5$  &$1.5$    &$0.5$ &$1.5$    &$0.5$ &$1.5$
% &$0.5$ &$1.5$&$0.5$ &$1.5$
\\ \hline
$ \alpha_s $    &$0$    &$0$    &$0$    &$0$    &$-0.3$    &$0.3$ &$-0.3$ &$0.3$
% &$-0.3$ &$0.3$&$-0.3$ &$0.3$
\\ \hline

$ \Omega_k $ &$0$    &$0$    &$0$    &$0$    &$0$    &$0$ &$0.05$ &$0.05$
% &$-0.05$ &$-0.05$&$0.05$ &$0.05$
\\ \hline

$ z_{re} $  &$4$    &$20$   &$4$    &$20$   &$4$ &$20$   &$4$ &$20$
% &$4$ &$20$ &$4$ &$20$
\\ \hline

$10^{10} A_s $ &$10$   &$100$  &$10$   &$100$  &$10$   &$100$  &$10$ &$100$
% &$10$ &$100$&$10$ &$100$
\\ \hline
\end{tabular}

\caption{Priors for  different parameters in the COSMOMC Runs.
Here $\Omega_b h_{\mt{out}}^2$ is the physical baryon density, $\Omega_{\rm m } h_{\mt{out}}^2$ is the physical dark matter
density, $z_{re}$ is the redshift at re-ionization, $n_s$ is the spectral index, $\al_s$ is the
running of the spectral index and $A_s$ is the amplitude of scalar fluctuations (for definitions see, {\it e.g.}~\cite{cosmomc}).}
\end{table}
We used the version of the COSMOMC program which lets one analyze the range $2\leq l\leq 30$ for TT
correlations  and the range $2\leq l\leq 23$ for TE+EE correlations using the
pixel-based approach (T, Q and U maps), which offers  a much more accurate treatment of the low-$l$ likelihood \cite{wmap3}.  One has
(957+1172) pixel data in all. The rest of the correlations that
we considered consisted of $C_l^{TT}$ in the range $31\leq l\leq 1000$, and  $C_l^{TE}$ in the
range $24\leq l\leq 450$.

\begin{figure}
\centering
\includegraphics[keepaspectratio=true,scale=.7]{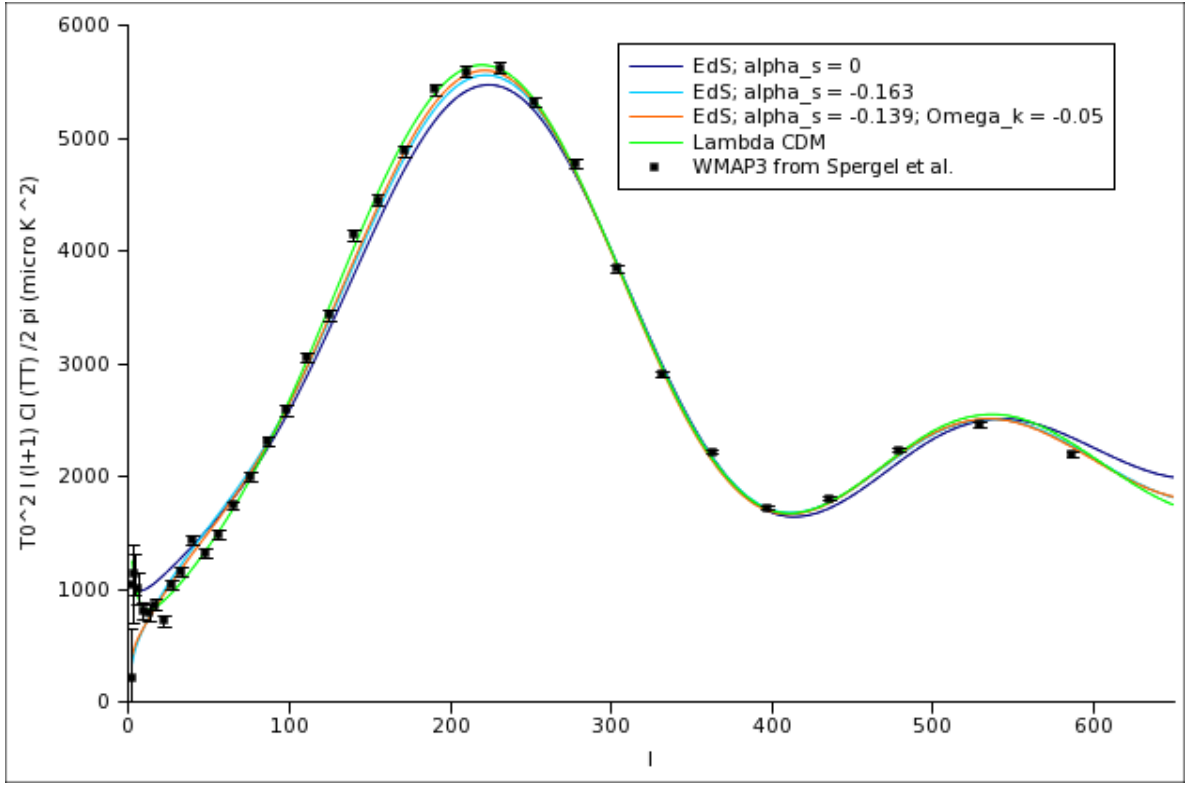}
\caption{$\La$CDM and EdS fits to the WMAP 3 binned data} \label{fig:TT_binned}
\end{figure}
\begin{figure}
\centering
\includegraphics[keepaspectratio=true,scale=.48]{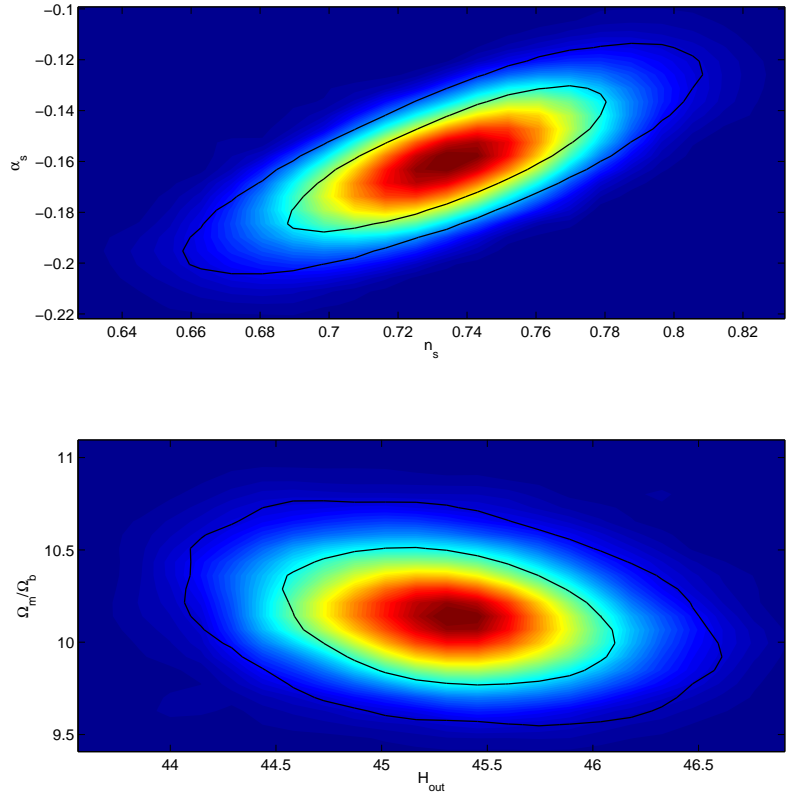}
\caption{Contour marginalized likelihood plots to the WMAP 3-yr data  for the run ``EdS, $\al_s\neq 0$''. The coloured map corresponds to mean likelihood, while the solid lines correspond to marginalized 1-$\s$ and 2-$\s$ contours.}
\label{fig:likelihood}
\end{figure}

We find that an EdS universe with {\it no dark energy}  but with a value of the Hubble constant,
$H_{\rm out}$, significantly lower than the conventionally accepted value of ~ 70 km/s/Mpc gives a
very reasonable fit to the CMB spectrum, see fig.~\ref{fig:TT_binned}.  For  the high multipoles ($31\leq l\leq 1000$) TT power spectrum
 our goodness-of-fit (G.F.) is around 2\%, compared to around $5\%$ of the concordant $\La$CDM model. For the overall fit of
both the TT+TE+EE spectrum involving 3520 d.o.f., the EdS model has a reduced\footnote{The ``effective'' $\chi^2$ is obtained directly from COSMOMC~\cite{cosmomc}. To obtain the reduced effective chi-square, $\chi^2_{\mt{eff},r}$, we just divide it with the number of independent degrees of freedom.}    $\chi_{\mt{eff,r}}^2$ of
1.016 with a 26\% G.F., as compared to the ``concordance'' $\La$CDM
model\footnote{The ``concordance'' best fit $\La$CDM model is obtained by running the COSMOMC program
including both the WMAP 3-yr and supernovae data. The best-fit $\La$CDM parameters for the  WMAP
3-yr data alone yield very similar  $\chi_{\mt{eff,r}}^2$, indicating the presence of  the well-known degeneracy in $\Om_M-h$ plane of
 the WMAP data. In fact, it is this degeneracy that we exploit to fit CMB with $\Om_M=1$ and low Hubble parameter.} with  $\chi_{\mt{eff,r}}^2=1.005$ and G.F.=41\% (see table~\ref{table:fits} for more
details).  The most likely parameter set along with their 1$\s$ bounds are tabulated in
table~\ref{table:parameters}; also see  the likelihood plots,
fig.~\ref{fig:likelihood}.

\begin{table} \centering
\begin{tabular}{|c|c|c|c|c|}
\hline
& $\Lambda$CDM & EdS, $\alpha_s = 0$ & Eds, $\alpha_s \neq 0$ & Eds,
 $\alpha_s, \Omega_k \neq 0$  \\ \hline
$ \Omega_b h_{\mt{out}}^2$ & $ 0.022_{-0.002}^{+0.002} $ & $ 0.022_{-0.001}^{+0.001}
 $ & $ 0.018_{-0.002}^{+0.001} $ & $ 0.019_{-0.001}^{+0.002} $ \\
  \hline
$ \Omega_m h_{\mt{out}}^2$ & $ 0.106_{-0.013}^{+0.021} $ & $ 0.198_{-0.011}^{+0.008}
 $ & $ 0.186_{-0.009}^{+0.011} $ & $ 0.167_{-0.007}^{+0.009} $ \\
  \hline
$ \Omega_{\Lambda} $ & $ 0.759_{-0.103}^{+0.041} $ & $ 0 $ & $ 0 $ & $0
  $ \\  \hline
$ z_{re} $ & $ 11.734_{-7.619}^{+4.993} $ & $ 8.697_{-6.694}^{+4.351} $
 & $ 13.754_{-5.752}^{+2.246} $ & $ 13.342_{-5.011}^{+2.55} $ \\
  \hline
$ \Omega_k $ & $ 0 $ & $ 0 $ & $ 0 $ & $ 0.05 $ \\  \hline
$ n_s $ & $ 0.96_{-0.04}^{+0.04} $ & $ 0.94_{-0.038}^{+0.021} $ & $
 0.732_{-0.071}^{+0.07} $ & $ 0.761_{-0.069}^{+0.069} $ \\  \hline
$ \alpha_s $ & $ 0 $ & $ 0 $ & $ -0.161_{-0.044}^{+0.044} $ & $
 -0.13_{-0.048}^{+0.037} $ \\  \hline
$ 10^{10} A_s $ & $ 20.841_{-3.442}^{+3.116} $ & $
 25.459_{-2.766}^{+2.135} $ & $ 25.302_{-2.968}^{+2.182} $ & $ 23.975_{-2.448}^{+2.198} $ \\
  \hline
$ \Omega_m/\Omega_b $ & $ 4.73_{-0.485}^{+0.999} $ & $
 9.119_{-0.357}^{+0.341} $ & $ 10.094_{-0.489}^{+0.645} $ & $ 8.929_{-0.541}^{+0.512} $
 \\  \hline
$ h_{out} $ & $ .72857_{-.07393}^{+.05137} $ & $ .46857_{-.01307}^{+.00888}
 $ & $ .4523_{-.01129}^{+.01291} $ & $ .42069_{-.00919}^{+.01107} $ \\
  \hline
$ Age/GYr $ & $ 13.733_{-0.369}^{+0.389} $ & $ 13.908_{-0.258}^{+0.399}
 $ & $ 14.408_{-0.4}^{+0.369} $ & $ 15.338_{-0.393}^{+0.342} $ \\
  \hline
$ \sigma_8 $ & $ 0.77_{-0.109}^{+0.121} $ & $ 1.012_{-0.081}^{+0.056} $
 & $ 0.919_{-0.075}^{+0.07} $ & $ 0.862_{-0.063}^{+0.06} $ \\  \hline
$ \tau $ & $ 0.095_{-0.074}^{+0.072} $ & $ 0.047_{-0.041}^{+0.037} $ &
 $ 0.079_{-0.044}^{+0.023} $ & $ 0.081_{-0.041}^{+0.024} $ \\  \hline
\end{tabular}
\caption{Most likely parameter values with 1 $\sigma$ errors for the various COSMOMC Runs}
\label{table:parameters}
\end{table}

\begin{figure}
\centering
\includegraphics[keepaspectratio=true,scale=.6]{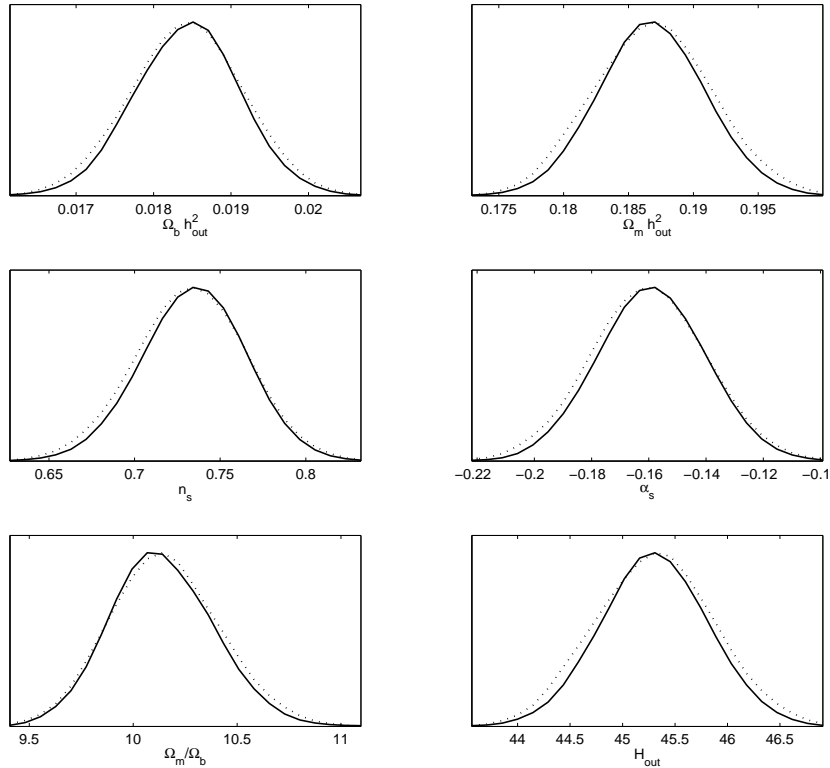}
\caption{Parameter likelihoods  to the WMAP 3-yr data for the run ``EdS, $\al_s\neq 0$''. Dotted lines are ``mean likelihoods'' of samples, while solid lines are ``marginalized probabilities''~\cite{cosmomc}.} \label{fig:1dlike}
\end{figure}
 We also produce two 2-dimensional likelihood contour plots: (i) $h_{\rm out}$ vs. $\Om_m/\Om_b$ which are the only two independent parameters related to the composition of the universe, and (ii) $n_s$ vs. $\al_s$ which characterize the spectrum.

\begin{figure}
\includegraphics[width=0.48\textwidth,angle=-90,scale=0.85]{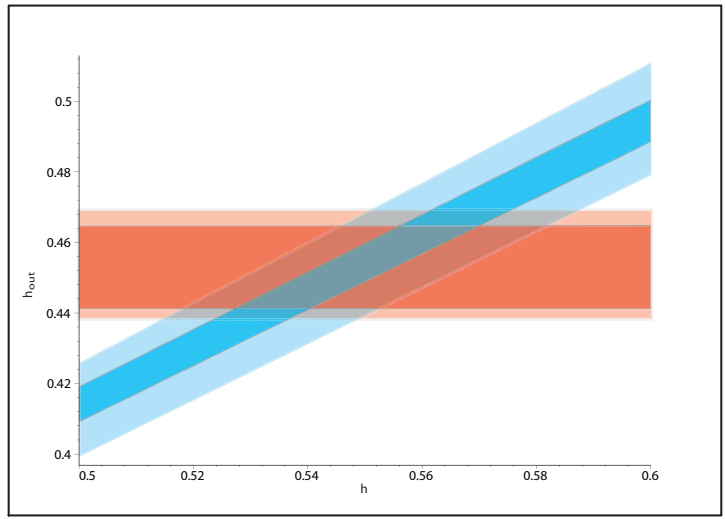}
\caption{ \label{fig:wmap-sn} 1-$\s$ and 2-$\s$ Contour plots for $h$ vs. $h_{out}$. The blue bands come from the SN-I analysis, while the red bands correspond to constraints coming from WMAP. }
\end{figure}

The most crucial quantity to consider is the Hubble parameter and in particular what a consistency
with the supernova data implies for the locally measured value. In fig.~\ref{fig:wmap-sn} we show a
contour plot combining the constraint from supernova fit in the previous section with that of
WMAP. As promised before, we find that the locally measured Hubble
parameter can be as high as $h\sim .59$ at the 2$\s$, or 95\% C.L., which is within the
acceptable range of the different measurements of the Hubble parameter.

Let us briefly discuss  about the values that we obtain for the other cosmological parameters, a
more detailed discussion on some of these constraints is presented in the next section. The main
constraint on the baryon density comes from BBN, and we are indeed consistent with the data (see
next section for details). As one can see from the likelihood plot, fig.~\ref{fig:likelihood}  as
well as table~\ref{table:parameters}, the ratio between dark matter and baryons  is somewhat
higher,  $\Om_m/\Om_b\sim 10$,  than the ``concordance'' $\La$CDM model value of  $\Om_m/\Om_b\sim 6$.
Measurements of light-to-mass functions in galaxy clusters can in principle be used to constrain
these numbers, but presently they suffer from  relatively large uncertainties (see for instance~\cite{light2mass}, and references therein). The issue is further
compounded by the fact that our local ratio of abundances within the LTB patch may not represent
the global ratio. A  more detailed investigation will be required to settle the issue, but
potentially this could be a problem. For the total matter density, one now has tight constraints
from the observation of BAO~\cite{Eisenstein}. As we discuss in the next section, the total matter
density in our model (which is the same as the critical density and hence $\propto h_{\rm out}^2$)
seems consistent with these measurements.

What about the properties of the primordial spectrum? Our best fit spectral tilt is relatively low, $n_s\sim .73$, but there are several inflationary scenarios where such low spectral tilts are common (for example in modifications of the old inflationary scenario from false vacuum~\cite{old}, or inflation from exponential potentials  naturally occurring in string theories, see for instance~\cite{exponential}). Our model also requires a significant running,  $\al_s\sim -.16$. It is a known fact that the 3rd year WMAP data favors a significant running of the spectral
 index which deviates from a Harrison-Zeldovich scale invariant scalar
 power spectrum.  For example, the analysis of~\cite{casas}  gives a running
 $\al_s = -0.055^{+0.028}_{-0.029} $ at $60 \% $ confidence level.  In
 fact, most inflationary models predict a running spectral index~\cite{LR} (see also~\cite{running}; models of inflation from a False Vacuum have typically an abrupt transition in the spectral index~\cite{old}) .
%Nonetheless,  for more exotic inflationary models,
% running is currently unconstrained.   Furthermore, when working without
% the cosmological constant as a prior, the running of the spectral index is completely
% uncostrained.
Additional constraints on
$\{n_s,\al_s,\s_8\}$ can mostly come from observations of large scale structure and weak-lensing
experiments. In the context of our MV model, this is a difficult and somewhat tricky task which we
have postponed to a future analysis, however we do discuss briefly possible implications in the
next section.

Finally, we note that our value of the re-ionization epoch (optical depth) is broadly consistent
with the usual observations~\cite{reionization} (see also discussion in~\cite{wmap3}).

To summarize, our best fit (WMAP + SNIa) MV model consists of 8 parameters, one of which, the length
scale of the void, has been chosen at the value $L = 250/h$ to derive our best-fit model. However, as noted in the introduction, if one  ``accepts'' a G.F. $\sim 10\%$ to the supernovae data, then one can go down to a much smaller length scale, $L\sim 160/h$. Out of the other seven parameters,  six of them (columns 2 to 7 in the  Table of \ref{table:mv1}  are obtained from the fit to the
WMAP 3-yr data using COSMOMC, while the last one, (column 8), is constrained from the supernovae data. We note that a ``minimally acceptable'' model with respect to the central underdensity contrast would be obtained with a maximally acceptable $h_{\rm out}\sim 0.47$, at the 95\% C.L.. This in conjunction with Eq.(\ref{local-h}), then tells us that the minimal jump parameter has to be $1.17$, or equivalently $\de_0\sim 0.44$.  Using these information we tabulate all the parameters in  Table \ref{table:mv1} for our ``best-fit'' and ``minimally-acceptable'' model. We note that the values of $\de_0$ and $L$ in the ``minimally-acceptable'' fit is not far from what observationally is suggested in~\cite{hole}.
% $\de_0\sim - 0.3$ and $L\sim 150/h$.
\begin{table}
\begin{tabular}{|c|c|c|c|c|c|c|c|c|c|c|}
\hline
Parameter& $L$&$\Om_b h_{\rm out}^2$ & $\Om_m h_{\rm out}^2$   & $z_{re}$   & $\s_8$  & $n_s$ & $\al_s$    & $\de_0$ &$h_{\rm out}$&$h$  \\ \hline

Best-fit & $250/h$&$ 0.018_{-0.002}^{+0.002}$ & $ 0.19_{-0.01}^{+0.01}$ & $13.8_{-5.8}^{+2.2}$ & $0.92_{-0.08}^{+0.07}$ &  $0.73_{-0.07}^{+0.07}$ &$ -0.16_{-0.04}^{+0.05}$ &  $0.51^{+0.03}_{-0.04}$ &$0.452^{+.013}_{-.011}$&$0.55^{+.024}_{-0.023}$\\ \hline
Acceptable-fit & $160/h$&$ 0.02$ & $ 0.2$ & $13.8$ & $0.92$  &  $0.73$ &$ -0.16$ &  $0.44$ &$0.47$&$0.55$\\ \hline
\end{tabular}
\caption{\label{table:mv1}Best-fit Minimal Void Model Parameters}
\end{table}

%%%%%%%%%%%%%%%%%%%%%%%%%%%%%%%%%%%%%%%%%%%%%%%
\section{Can we improve WMAP and supernovae fits?}
We have seen that by allowing significant running in the range of the observed CMB spectrum one is
able to obtain a reasonable fit to the WMAP 3yr data. However, the overall fit is not as good as the
best-fit $\La$CDM model. Secondly, as is clear from the combined contour plot
fig.~\ref{fig:wmap-sn}, consistency with WMAP and supernovae data requires a relatively low local
value of the Hubble parameter. The underdensity contrast required is also quite high (centrally around 50\%, and on average around 35\% in the Void). Can we somehow modify the MV model to get a better fit and overcome these difficulties?
%Also, the spectral tilt required is somewhat low, and one also requires a relatively large running (and one still would need to perform a quantitative check with the matter power-spectrum from large scale structure measurements, which we postpone to future work). Can we get parameters which are closer to the common lore?
We now discuss two different modifications in this context.
\subsection{``Bump'' Model} \label{Bump}
The first one concerns using  different ``priors'' for the primordial spectrum. For instance,
in~\cite{sarkar2} the authors  assumed the existence of a bump in the primordial spectrum as a
prior, rather than considering an overall running as we do, in order to fit the CMB data without Dark Energy. Although in these models the number of
parameters is larger than what we consider, one obtains much better fits to the WMAP data (in fact,
slightly better than $\La$CDM), and is thus worth investigating further. Such a bump can be
produced by a rapid succession of two phase transitions~\cite{sarkar2} and is thus
phenomenologically well-motivated. Moreover, it is  rather intriguing and promising to note that~\cite{sarkar}
such  a bump  would also enhance the probability of having voids today at the scale of the bump itself,
which happens to be approximately the same scale we are considering here. This ``bump'' model, in its original form, of course cannot reproduce the supernovae data, and the Hubble
parameter ($h_{\rm out}\sim0.44$) is too low. So it seems natural to merge this model with our MV scenario.
Can the parameter set obtained be
consistent with the supernovae analysis that we have performed using the local void?

\begin{figure}
\includegraphics[width=0.48\textwidth,angle=-90,scale=0.8]{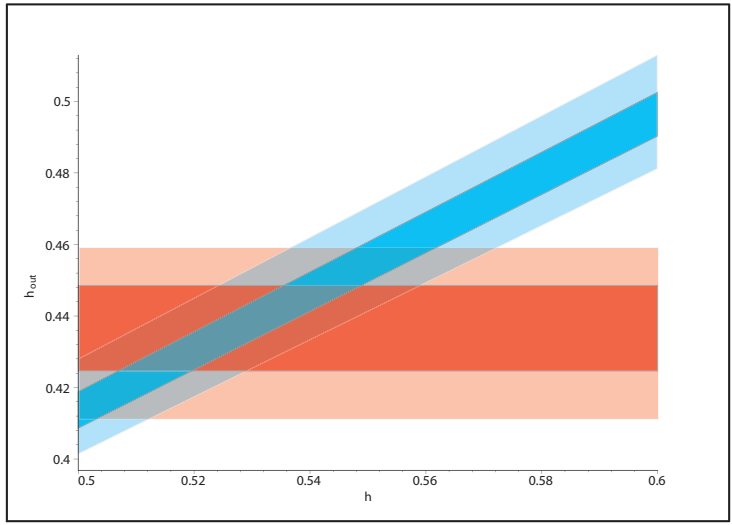}
\caption{ \label{fig:sarkar}  1-$\s$ and 2-$\s$ Contour plots for $h$ vs. $h_{out}$ for CDHM bump model~\cite{sarkar}. The blue bands come from the SN-I analysis, while the red bands correspond to constraints coming from WMAP.}
\end{figure}

Of course, having a local void again ensures that  the supernovae data is consistent. The crucial question is
whether putting together the MV framework with the ``bump'' model  could lead to an ``acceptable'' local Hubble
parameter.  As we see in the contour plot (see fig.~\ref{fig:sarkar}) at the 95\% C.L. one can have
as high as $h\sim .57$, which is definitely within the acceptable range Eq.(\ref{local-h}).

\subsection{Adding Curvature}  \label{curvatura}
In this subsection we consider a different possibility, namely adding  curvature to the
model. Although having curvature would be considered fine-tuning in an inflationary paradigm, we
point out that the low multipole anomalies~\cite{lowl}, if taken seriously, could be suggestive of having only the
``minimal'' number (50-60 depending upon the reheating temperature) of efoldings, which would be
consistent with having a slight curvature. Also, we note that other models involving cyclic
scenarios typically do not predict a flat universe to any high precision.

Accordingly, we  performed a run where we allowed up to 5\% in curvature  along with  including  running of
the tilt, as before. We found that the best-fit parameter set prefers the highest value of spatial curvature that we allowed. Consequently, we performed a run with $\Om_k=0.05$, corresponding to a slightly closed universe to see how curvature may affect the goodness of fit\footnote{We are currently pursuing a more exhaustive analysis of the void model with curvature.}. We now indeed find a much better fit to the WMAP data. For the overall TT+TE+EE
data, $\chi^2_{\mt{eff,r}}=1.012$ corresponding to a 31\% goodness-of-fit (see
table~\ref{table:fits} for more details). The Hubble parameter, however is slightly lower than our
previous results, as can be seen from the likelihood plots involving (ii) $\Om_k$ and $h_{\rm
out}$, also see table~\ref{table:parameters}. None of these results are very surprising or new. Previous studies had already observed that one can get good fits to WMAP with a closed universe, but it is precisely because of the rather low value of the Hubble parameter required for these fits that these models are not considered seriously. However, when combined with the jump
parameter obtained from the supernovae analysis\footnote{In principle once one adds curvature, one
has to redo the analysis of the supernovae data set. We have not done it for this preliminary
analysis, because we do not expect any significant difference from the small amount of curvature
that we  allow. } given by~Eq.(\ref{SNJump}), this gives us a local Hubble parameter which can be
consistent with observations, given in~Eq.(\ref{local-h}).

\begin{figure}
\centering
\includegraphics[keepaspectratio=true,scale=.6]{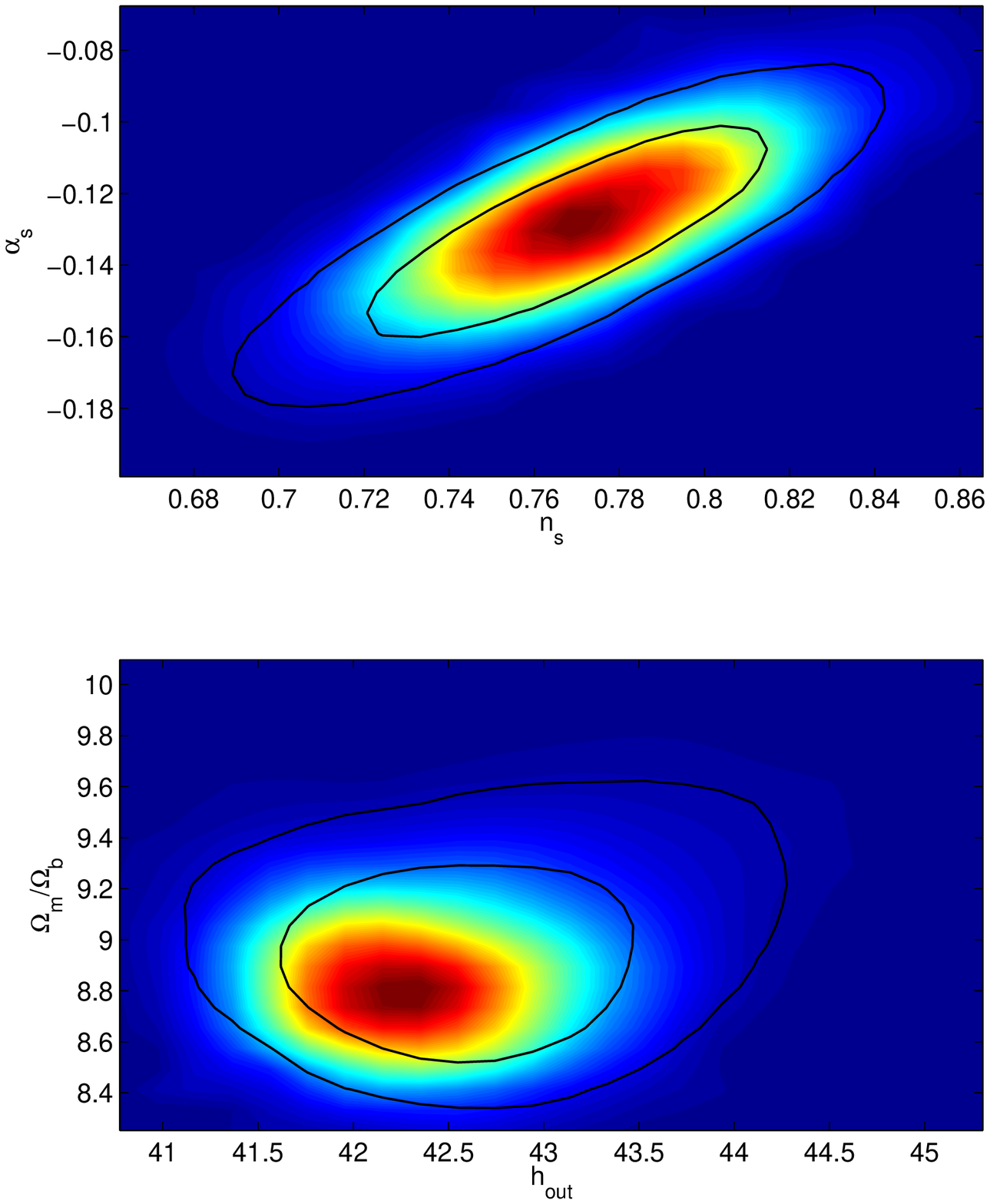}
\caption{Marginalized Likelihood plots for the WMAP 3-yr data for the run ``EdS, $\al_s,\ \Om_k\neq 0$''.  The coloured map corresponds to mean likelihood, while the solid lines correspond to marginalized 1-$\s$ and 2-$\s$ contours.} \label{fig:curv-likely}
\end{figure}

It is also worth pointing out that, as is clear from the contour plot fig.~\ref{fig:curv-likely},  there is a degeneracy direction in the WMAP data where as we simultaneously increase the curvature and the Hubble parameter we can still get good fits. This suggests that even if we allow for a slightly closed universe, by decreasing the Hubble parameter slightly (from our EdS value) we may be able to get significantly better fits to the WMAP data.  In other words, when combined with other data, such as measurements of local Hubble
parameter which  prefer higher values of the Hubble parameter, the MV model may still
provide a reasonable fit.

In passing we note that, the 2$\s$ range
for the tilt and the running is much closer to the conventional values as compared to our
original MV model (see figs.~\ref{fig:likelihood} and \ref{fig:curv-likely}, for comparison).
\begin{table}
\begin{tabular}{|c|c|c|c|c|c|c|}
\hline
 & \multicolumn{2}{c|}{\bf{$C_l^{TT}$}}& \multicolumn{2}{c|}{\bf{$C_l^{TT}+C_l^{TE}$}} & \multicolumn{2}{c|}{\bf{Total}}\\
\hline

Model & $\chi^2_{eff}$   & G.F.   & $\chi^2_{eff}$    & G.F. & $\chi^2_{eff}$    & G.F. \\ \hline

Concordant $\La$CDM & 1038.9 & 4.7\% & 1455.2 & 11.3\% & 3538.6 & 41\%  \\ \hline

EdS $\alpha_s=0$ & 1124.6 & 0\% & 1711.9 & 0\% & 3652.3 & 6\%  \\ \hline

EdS $\alpha_s\neq 0$ & 1057.8 & 1.9 \% & 1475.5 & 5.7\% & 3577.4 & 24.6\%  \\ \hline

EdS $\alpha_s, \ \Omega_k\neq 0$ & 1048.7 & 2.9\% & 1466 & 7.9\% & 3560.9 & 31.1\%  \\
\hline
\end{tabular}
\caption{\label{table:fits} $\chi^2_{eff}$ and goodness-of-fit for the different COSMOMC Runs. The first column corresponds to high-$l$ TT power spectrum, ($31\leq l\leq1000$). The second column corresponds to both the high-$l$ TT ($31\leq l\leq1000$) and TE ($24\leq l\leq 450$) data. Finally, the last column contains the total statistics of  TT ($2\leq l\leq1000$) and  TE ($2\leq l\leq 450$) spectrum.  }
\end{table}

%%%%%%%%%%%%%%%%%%%%
\section{Consistency with Other Observations}\label{other}
Here we briefly discuss the consistency of our MV model with observations other than the supernovae,
WMAP and local measurements of  the Hubble parameter.

\noindent
{\bf BBN:} Primordial Nucleosynthesis has been a spectacular success story for the Standard Big Bang paradigm which predicts specific freeze out abundances of light elements such as $D$, $He^3$, $He^4$ and $Li$. These freeze out abundances depend on the baryon-to-photon ratio. Since from the measurements of CMB temperature we know the photon energy density precisely, BBN can also constrain the baryon density in our universe today. The success of the BBN paradigm lies in the general agreement of this number, measuring the abundances of the different light elements spanning 9 orders of magnitude (for a review see~\cite{sarkarbbn}). BBN therefore constraints the baryon density, so that at 95\% C.L. we should have\footnote{While estimates from $He^4$ and $D$ are slightly higher, measurements using $Li$ suggests a lower number.} $0.017\leq \Om_b h_{\rm out}^2\leq 0.024$. This is indeed consistent with the parameter range that we obtain from the WMAP run, $\Om_b h_{\rm out}^2=0.018^{+.002}_{-.002}$. It is remarkable that although we have a higher baryonic abundance, the lower Hubble parameter almost precisely compensates to yield approximately the same baryonic energy density as it is obtained in the ``concordance'' $\La$CDM model.\\
%{\bf Matter density from Clusters:}
%Observations from rich galaxy clusters can provide us with the ``light-to-mass'' ratio in these clusters. While the amount of light is proportional to the baryon density, the mass contains both the dark and baryonic matter. Although this ratio seems to vary with the scale of the mass distribution and in general is different from the cosmic mean ratio, one can estimate $\Om_d/\Om_b$ from these sort of measurements, with possibly a 15\% to 20\% overall uncertainty. {\it i am still trying to find a good review with numbers.}\\

\noindent {\bf BAO:} Recently, a remarkable achievement of observational cosmology has been to
identify the baryon acoustic peak in the galaxy-galaxy correlation function using Luminous Red
Galaxies (LRG's)~\cite{Eisenstein}. The overall shape of the galaxy correlation function mainly depends
on the shape of the primordial spectrum (tilt and running) and the epoch of matter-radiation
equality (scales which entered the Hubble horizon before the equality have their amplitudes
relatively suppressed as compared to the ones which entered later). On top of the ``overall
envelope'' one has now observed a tiny peak coming from the baryon acoustic oscillations. The
position of the peak is related to the sound horizon of the baryon-photon plasma at the time of
recombination. In fact, what one really measures is more like an angle which is the ratio of the
sound horizon at recombination (evolved at $z\approx 0.35$, which is the average redshift of the LRG survey) and the angular
distance\footnote{In the survey, one really measures a combination of the angular and transverse
distances, see~\cite{Eisenstein} for details.} at the same redshift $z\sim 0.35$. This ratio therefore is not
only sensitive to the baryon density in the universe, but also to the evolution of our late-time
universe and therefore, to the amount of dark energy, for instance. Using essentially the two pieces
of information (overall shape and peak) one is able to constrain {\it two} different quantities,
for instance, the matter density and $d_V$ (a specific combination of the transverse and angular
distance at $z\approx 0.35$~\cite{Eisenstein}). This in turn can constrain the composition of the universe and it was
claimed in~\cite{Eisenstein} that a pure EdS model is ruled out at the level of 5$\s$. Can the MV
model  be consistent?

Firstly, it is difficult to provide a crisp answer to that question based on the analysis done
in~\cite{Eisenstein} because the analysis of the data (conversion from redshift to distance etc.)
is done using the ``concordance'' $\La$CDM model. In particular we point out that precisely in the
redshift range of the sample, $0.16<z<0.47$, the luminosity distance vs. redshift curve of the void
model (which is the same as an EdS model, in this range)
differs significantly from the $\La$CDM curve. Thus to be precise, one needs to reanalyze the
LRG data in the context of an EdS model. Nevertheless,
one can try to see whether one can satisfy the bounds on $d_V$ and $\Om_m h^2$ that was placed
in~\cite{Eisenstein}: \be d_V=1370\pm 128 \, \qquad \mx{ and }\qquad \Om_m h^2=0.130 \, (n_s/0.98)^{1.2}\pm 0.022 \, , \ee where
the errors correspond to approximate 2$\s$ (95\% C.L.) values. Now, in our model, we have
a low spectral tilt (and also a relatively large running, which can  alter the shape of the
correlation function and hence the constraints). Relegating a more systematic analysis for future,
and just correcting for the lower tilt in our model implies the following constraint for the total
matter density, which is given by the average Hubble parameter in our model: \be \Om_m h^2\ra
h_{\rm out}^2=0.185\pm 0.022 \, , \ee where we have used our best-fit spectral tilt, $n_s\approx
0.73$. We recall that  our best fit Hubble parameter gives $h_{\rm out}^2\sim 0.205$, and therefore
is consistent with the above bound.

On the other hand the angular distance at $z=0.35$ for our model does not appear to be consistent with the values reported in~\cite{Eisenstein}.
In fact one can check that an EdS model has roughly the same distance of a concordance $\Lambda$CDM model at $z=0.35$ if the ratio
of the value of the Hubble constants of the two models is around $1.2$.
Since the concordance model (which fits the BAO scale) has $h\approx 0.7$, an EdS that fits this scale should have $h_{\rm{out}}\approx 0.7/1.2\approx 0.58$.
As we have discussed, this value is too large with respect of our analysis of the WMAP data.
More work is needed in order to find whether it would be possible to overcome this potential problem: for example adding more curvature could give a higher $h_{\rm out}$, from WMAP (which, by the way, would make the whole scenario in better agreement with other data as well: for example with the local measurements of $h$). It has to be seen, through a combined statistical analysis including the BAO data, whether this could give a consistent picture.
A different (though not very appealing) possibility, which would certainly work, is to make the Void much larger, extending up to redshifts of order $z\approx 0.4$.
We have checked that this can give the correct distance at $z\approx 0.35$ (see also the recent analysis in~\cite{Troels}), and that moreover this gives also a good fit of the BAO scale at $z\approx 0.2$, given in~\cite{Percival}.

\noindent {\bf Observations from Large scale Structure and Weak lensing:} An important
class of cosmological observations comes from large scale structures and weak lensing observations.
These typically produce constraints on $\s_8$, as well as on the shape of the primordial spectrum, $n_s$ and $\al_s$ (for instance
using Lyman-$\al$ forest). However, as mentioned above, one has to revisit these analysis in the
light of MV model, as one has a non-standard $D_L(z)$ relation.
We leave for future such a careful study of the large scale structure, Lyman-$\alpha$ and weak
lensing  data. Let us still make a few brief comments.

About $\s_8$,  at first sight our value is a bit high, $\s_8=0.92^{+.07}_{-.08}$, in our model, but even if this
situation turns out to be incompatible with the large-scale structure data (after a careful study), this may only be indicative of the need to include some
hot dark matter component~\cite{sarkar,sarkar2}. In the light of neutrinos having mass, this is a
perfectly natural scenario to consider. About $n_s$ and $|\al_s|$, the values are respectively
lower and higher than what the conventional $\La$CDM analysis suggests and in particular one may
worry about conflicts with Lyman-$\al$ measurements. However, we firstly point out that an analysis
of the Lyman-$\al$ measurements has to be now re-done in combination with the different set of
priors that we use to study the WMAP data. Secondly, introducing new physics, such as including a
little curvature, can push the values of  $n_s$ and $\al_s$  much closer to the standard values. In
short, there are too many uncertainties for us to make here any concrete conclusions, and one really needs
to perform a careful study of the above mentioned observations.

\noindent {\bf ISW Correlations}: Another interesting piece of evidence for Dark energy is given by
the Integrated Sachs Wolfe effect, which is claimed to be detected with some significance by some collaborations~\cite{isw}.
The detection is a correlation between the CMB maps of the sky and the galaxy surveys, which cannot
be explained in an EdS universe (since in this case the
linear gravitational potential does not evolve and therefore CMB photons do not get any net
frequency shift when passing through a potential well), and therefore are interpreted as
independent evidence for Dark Energy (since the potentials can evolve in $\Lambda$CDM). However the effect is absent only at the linear level, and it
exists also in EdS in the presence of nonlinear gravitational clustering. This is usually assumed
to be smaller then $10^{-5}$, but it actually happens to be of order $10^{-5}$ (and thus, visible
in the CMB) for structures as large as those that we are proposing in the present paper (few
hundreds of Mpc). It would be interesting to try and reproduce the ISW detection assuming the presence
of large voids and structures in the sky.

Moreover in the local underdense region we have assumed that the growth of fluctuations is different than the flat CDM model (it is in fact more similar to an open Universe): this leads also to an ISW effect for density fluctuations localized inside the Void. Studying this effect would be very interesting and could significantly affect the low-$l$ part of the CMB spectrum and therefore also the parameter estimation from the CMB. However this goes beyond the scope of the present paper since it would require a full treatment of the growth of density fluctuations in an LTB metric (this problem has been recently attacked by~\cite{Zibin}).

%%%%%%%%%%%%%%%%%%%
\section{Conclusion and Discussion}

The Type Ia supernovae data reveal that our universe is accelerating at redshifts that approximately correspond to the epoch of non-linear structure formation on large scales (the epoch of the formation of the so-called ``cosmic web'').
Given this fact, we have explored the possibility that the effect of
a large scale void can account for this acceleration due to
a jump between the local and the average Hubble parameter, instead of invoking a spatially constant dark
energy/cosmological constant component.  We find that the Minimal Void (MV) model can consistently account
for the  combination of the Type Ia supernovae, WMAP 3rd year, BBN
constraints, provided that the void spans a radius of about of $200\ {\rm Mpc}/h$ with a relative under density of
$45 \%$, near the center.  The MV model can accommodate reasonably all of the data
considered, although the fits are not as good as the concordance model.
However, we see the possibility of
obtaining just as good fits when one includes curvature or invokes non-standard features in the primordial spectrum (a ``bump'' for example).
We leave these issues for an upcoming work.
On the other hand we have seen that the Minimal Void is in trouble with the Baryon Acoustic Oscillations measurements, since outside the Void, the
$D_L(z)$ curve is just the usual EdS one, and the Hubble parameter $h_{\rm out}$ from WMAP is too low. More work is needed in order to find whether it would be possible to overcome this potential problem (for example by finding a fit for WMAP with  higher $h_{\rm out}$).

We end with observational and theoretical possibilities of distinguishing the MV model from $\Lambda CDM$.  The MV model predicts that the spectral index has to run significantly in the
WMAP3 data and that the ``average'' Hubble constant ({\it i.e.} outside the local region) has to be around $h_{\rm out} \sim 0.45$.   The $\Lambda CDM$ model, instead, requires a finely tuned
cosmological constant or dark energy component, in order to be consistent with the same data set.  Both cases
require significant model building and new physics that are currently being pursued by the
community.  How are we to distinguish between these two models?  The first logical way seems to perform  galaxy counts up to very large distances and in a wide area in the sky, in order to directly check if we could really live inside a huge Void. Moreover, there are features which can be checked by looking at SN Ia themselves: firstly,  the luminosity-redshift
curve in the two models deviate from each other significantly at redshifts $z \geq 1$. Secondly,   in the
MV model the curve  has a sharp peak (in correspondence with the boundary of the local region) around $z\simeq 0.1$,
while this peak does not exist in the $\Lambda CDM$ model.  The up-coming experiment SDSS-II~\cite{sdss2} will probably be able to discriminate the presence of such a peak.
Another unique prediction for the MV model comes from realizing that
the void is not expected to be exactly spherically symmetric, which could lead to detectable
anisotropies in the Hubble parameter as well as in the low multipoles in CMB. Additionally, these
anisotropies should be correlated!
We note, also, that one could be able to constrain Voids by looking at the blackbody nature of the CMB~\cite{geocentrism,copernican}. Our MV is still consistent with these constraints (while, according to~\cite{copernican}, voids that extend up to $z\sim 1$ are excluded).
Finally, studying large scale structure (as we plan to do in future work) one can study the compatibility of the primordial power spectrum we are assuming (with low tilt and large running, or with a bump) with the matter power spectrum. It may also be possible to test the existence of such a large running using Planck-satellite data as suggested by the Bayesian analysis performed in~\cite{liddle} using simulations.

In conclusion, we have shown that, for WMAP and SNIa observations, the
MV model could be taken as an alternative to invoking a dark energy component that will be
further tested in forthcoming supernovae observations.  On the other hand this has to be made consistent also with the Baryon Acoustic Oscillations. On the theoretical end, much work needs to
be done to establish if such large voids can actually be produced in our Universe by generic physics of
structure formation.  We are currently pursing this issues.
\vs\\
{\bf Note Added:} Most of the above research work was completed before the release of the WMAP 5yr data and we have decided not to re-analyze the CMB data in the present paper for the following reason: although the 5-yr data improves the 3-yr data, there is no significant qualitative difference between the results presented in the 3-yr and 5-yr survey. In this context, we further emphasize that our aim in this manuscript is not to compete with $\La$CDM on the basis of Bayesian likelihood analysis (in which case the  analysis can be very sensitive to the data, for instance  a difference of $\chi^2\sim 1-2$ may be significant), but to simply present a model which can be consistent with the data on the basis of the goodness of fit (for instance, a difference of $\chi^2\sim 1-2$ does not significantly reduce the goodness of fit). In addition a more systematic treatment including other cosmological data (BAO, Large Scale Structure data) and more recent data (CMB and Supernovae) is the subject of a future publication.

\acknowledgments We would like to thank  Robert Brandenberger, Paul Hunt, Subir Sarkar, Ravi Seth and  Tarun Souradeep, for
useful discussions and suggestions. We would specially like to thank Suman Bhattacharya, Jason Holmes and Antony Lewis for their help with COSMOMC. We thank Sebastian Szybka and Seshadri Nadathur for pointing out typos in earlier versions of the manuscript.

%%%%%%%%%%%%%%%%%%%%
%\appendix Appendix: Analytical Results for LTB Metric
\section{Appendices: Analytical Results for LTB Metric}\label{appendix}
\subsection{Metric \&  Density Profile}
In our paper we are interested in a special class of exact spherically symmetric solutions of
Einstein's equations with dust, known as the ``open''  LTB metric (in units $c=1$). We follow the treatment given in~(\cite{bmn,swiss}), where we have set the ``mass function'' to be cubic, which amounts to a redefinition of the radial coordinate (which is always possible if the mass function is a growing function of $r$). The metric is given by:
\begin{equation}
ds^2=-dt^2 + S^2(r,t)dr^2 + R^2(r,t)(d\theta^2 + \sin^2 \theta d\varphi^2) , \label{eq:14} \,
\end{equation} Here we have employed comoving coordinates ($r,\theta,\varphi$) and proper time $t$.
The functions $S^2(r,t)$ and the dust density $\rho(r,t)$ is given in terms of $R(r,t)$ via
\begin{eqnarray}
S^2(r,t) &=& {R^{'2}(r,t)\over {1+2(\bM r)^2k(r)}} , \label{eq:15}   \\
\rho (r,t) &=& {\bM^2M_{p}^2 r^2 \over R'(r,t) R^2(r,t)}  \label{eq:17} \, ,
\end{eqnarray}
where a dot denotes partial differentiation with respect to $t$ and a prime with respect to $r$,
while the function  $R(r,t)$ itself is given implicitly as a function of $(r,t)$ via an auxiliary
variable $u(r,t)$: \be R(r,t)={2\pi r\over 3 k(r)}(\cosh u-1)\, , \label{Ru}  \ee \be \tau^3\equiv
\bM t={\pi\stwo \over 3 k(r)^{3/2}}(\sinh u-u)\label{tu} \, , \ee In the above expressions, the
``curvature'' function $k(r)$ is left arbitrary (except that $k(r)\geq 0$) and this is what controls the density profile
inside the LTB patch, while $\bM$ is just an arbitrary (unphysical) mass scale.  Also,  we have
introduced the dimensionless conformal time $\tau$ for later convenience.

We also note that the average density inside the LTB patch is equal to the outside FLRW density~(see for instance~\cite{bmn,swiss}),
in the limit in which we can neglect $(\bM r)^2k(r)$ in~Eq.(\ref{eq:15}) in the spatial metric when defining the average (in our case the correction is always negligible).

To get an intuitive and analytical understanding of how the density profile is related to the
curvature function it is instructive to look at the ``small-$u$'' approximation where we only keep
next-to-leading terms in Eq.(\ref{Ru}) and Eq.(\ref{tu}). This gives us Eq.(\ref{matterdensity}).

%%%%%%%%%%%%%%%%%%%%%%%%%
\subsection{Photon Trajectories}
In order to perform supernovae fits we need to compute the luminosity (or angular) distances and
redshifts for a  photon trajectory  emanating (backwards in time)  from the central observer. The
first step in this direction is to solve for the photon trajectory: \be ds^2=0\Ra {dt(r)\over
dr}=-{R'(r,t(r))\over \sqrt{1+2(\bM r)^2k(r)}} \, . \label{tphoton} \ee The negative sign in front
takes care of the fact that the time increases as the photons go towards the center. Analytical
progress in solving the above equation is possible by realizing two things. Firstly, all quantities
($t(r),z(r),D_L(r)$) can be expressed as a power series in, $\bM r\sim r/R_H$, and since this is a
small quantity for the relevant inhomogeneous patches, we can just keep the next-to-leading order
terms in these expansions~\cite{swiss}. Secondly, formally one can combine  Eq.(\ref{Ru}) and
Eq.(\ref{tu}) to give us a power series expansion for $R(r,t)$ explicitly in terms of
$(r,t)$~\cite{swiss}:
\be R(r,t)=\3\pi \gamma^2r\tau^2\LF 1+R_2u_0^2+R_4u_0^4+\dots\RF \equiv \3\pi
\gamma^2r\tau^2\LF 1+f(u_0^2)\RF \label{R-u0} \, ,
\ee
where
\be u_0\equiv \gamma(\bM
t)^{1/3}\sqrt{k(r)}\mx{ and }\gamma\equiv \LF{9\sqrt{2}\over \pi}\RF^{1/3} \label{v} \, . \ee
It is important to realize that the
coefficients $\{R_n\}$, and hence the function $f$ are universal (do not depend on the specific
curvature function). It is implicitly defined via
\be
1+f(u_0^2)\equiv {2(\cosh u-1)\over u_0^2}\,\mx{\hs and\hs }6(\sinh u-u)=u_0^3
\label{f-defn}
\, . \ee
  This is what allows us analyze the problem in its full generality.

It is convenient to recast the equation in terms of the conformal time, $\tau$, and the
dimensionless radial coordinate \be \rb=\bM r \, . \ee Substituting Eq.(\ref{R-u0}) in Eq.(\ref{tphoton}) one
finds \be {d\tau\over d\rb}=-{{\pi\over 9}\gamma^2\LT1+\sum_1^{\infty}R_{2n} \gamma^{2n}\tau^{2n} (\rb
k^n)'\RT\over \sqrt{1+2k\rb^2}} \label{tau-series} \, . \ee The prime now denotes differentiation
with respect to the rescaled $\rb$. This can now be solved perturbative in $\rb$ to give us \be
\tau=\LF\tau_0 -{\pi\over 9}\gamma^2\rb\RF-{\pi\over 9}\gamma^2 \rb\sum_1^{\infty}R_{2n}
\gamma^{2n}\tau_0^{2n} k^n(\rb)+\cO(\rb^2) \, . \ee The first two terms within the brackets corresponds
to the FLRW expression for the trajectory while the rest of the terms give us the largest
corrections coming from the inhomogeneities within a local patch. For corrections outside the patch
see~\cite{swiss}. By comparing with Eq.(\ref{R-u0}) the above expression can succinctly be written as
\be \tau(\rb)=\tau_F(\rb)-{\pi\over 9}\gamma^2 \rb f(\gamma^{2}\tau_0^{2} k(\rb)) \, , \ee where the subscript
$F$ corresponds to FLRW.
%%%%%%%%%%%%%%%%%%%%%%%%%%%%%%%%%%%%%%%%%
\subsection{Luminosity Distance vs. Redshift}
Having found the photon trajectory, the next step is to compute the redshift which is governed by
the differential equation~\cite{celerier}
 \be {dz\over dr}= {(1+z)\dot{R}'\over \sqrt{1+2k\rb^2}}
\label{redshift} \, . \ee
 Again, if we are only interested in computing corrections up to linear
order in $\rb$, then the redshift is given by
\be \int {dz\over 1+z}\approx {2\pi\gamma^2\over
9}\int {d\rb\over\tau} [1+ \sum_n(n+1)R_n \gamma^{2n} \tau^{2n} (r k^n)'] ={2\pi\gamma^2\over 9}\int\LT
{d\rb\over \tau} +d\rb\sum_n(n+1)R_n \gamma^{2n} \tau^{2n-1} (r k^n)' \RT \, . \ee
To evaluate the first
integral we note that we can replace $\tau$ by $\tau_F$ as we will only be making an $\cO(\rb^2)$
error. Thus we have
$$\int{d\rb\over \tau}\approx \int{d\rb\over \tau_F}=-{9\over \pi\gamma^2}\int{d\tau_F\over \tau_F}=-{9\gam^2\over \pi}\ln{\tau_F\over \tau_0}$$
The second term can be integrated straight forwardly up to linear terms in $\rb$:
$$\sum_n(n+1)R_n \gamma^{2n} \int\tau^{2n-1} (r k^n)'d\rb\approx \sum_n(n+1)R_n  \gamma^{2n}\tau_0^{2n-1}\int(\rb k^n)'d\rb=\sum_n(n+1)R_n  \gamma^{2n}\tau_0^{2n-1}\rb k^n$$
$$=\rb [f(\gam^2\tau_0^2k(r))+\gam^2\tau_0^2k(r)f_1(\gam^2\tau_0^2k(r))]/\tau_0$$
where we have defined
\be
f_1(x)\equiv {df(x)\over dx}
\ee
Putting everything together we have
\be 1+z=\LT{\tau_0\over\tau_F(\rb)}\RT^2\exp\left\{{2\pi\gam^2\rb[f(\gam^2\tau_0^2k(r))+\gam^2\tau_0^2k(r)f_1(\gam^2\tau_0^2k(r))]
\over 9\tau_0}\right\} \, . \ee
Thus we have obtained an analytical approximation for the
redshift as a function of the radial coordinate. We note in passing that the term in front of the
exponential precisely correspond to the FLRW result. The corrections come from the exponential. In
fact for small $z$ one finds
\be z\approx{2\pi\over 9\tau_0}\gam^2\rb[1+f(\gam^2\tau_0^2k(r))+\gam^2\tau_0^2k(r)f_1(\gam^2\tau_0^2k(r))]
\label{redshift-soln} \, . \ee

The luminosity distance, in General Theory of Relativity, is related to the angular diameter
distance, $D_A$ via \be D_L=(1+z)^2D_A \, . \ee Now, in an LTB model when the observer is sitting at the
center, the angular distance is simply given by \be D_A=R=\3\pi \gam^2r\tau^2\LF
1+f(\gam^2\tau_0^2k(\rb))\RF \label{angular} \, . \ee Thus we now have both the luminosity distance and
the redshift as a function of the radial coordinate and one can easily plot $D_L(z)$ and check
whether the local void model can provide a good fit to the supernova data or not.

%%%%%%%%%%%%%%%%%%%%%%%%%
\subsection{The ``Jump''} \label{thejump}
A particularly important quantity that can be inferred from the $D_L(z)$ curve is the jump
parameter, $\cJ$ defined by Eq.(\ref{jump}).  Surprisingly, this turns out to not depend on the
specific profiles, let us here see this analytically. First observe that since $k'$ vanishes at
$r=0$, we have the general result \be R'(0,t)=\3\pi\gam^2\tau_0^2(1+f_0) \, , \ee where $f_0$ corresponds
to the value of $f$ at $r=0$. Then using the exact expression for the density function
Eq.(\ref{eq:17}) one finds \be \rho (r,t) = {M_{p}^2  \over 6\pi t_0^2(1+f_0)^3} \, . \ee
The underdensity
contrast at the center, $\de_0$ now can be easily related to $f_0$: \be
\de_0=(1+f_0)^{-3}-1\Ra1+f_0=(1+\de_0)^{-1/3}
\label{f-de}
\, . \ee

Now, on the other hand using the definition of the Hubble parameter Eq.(\ref{Hubble}), the correction
to the redshift Eq.(\ref{redshift-soln}), and the luminosity distance Eq.(\ref{angular}) one finds
$$H_0^{-1}=H_{\mt{out}}^{-1}{1+f_0\over 1+f_0+u_0^2f_{1,0}} \, . $$
Or in other words \be \cJ={h\over h_{\rm out}}={1+f_0+u_0^2f_{1,0}\over 1+f_0} \, . \ee
Since $\de_0$ uniquely determines $f_0$ via (\ref{f-de}), and $f(u_0^2)$ is a given function, it also determines $u_0^2$ and $f_{1,0}\equiv f_1(u_0^2)$. Thus in turn it also determines the jump parameter uniquely.

%%%%%%%%%%%%%%%%%%%%%%%%%%%%%
\subsection{CMB dipole moment}\label{dipole}
Let us consider our observer to be located slightly off-center, at $r=r_O$. In this case the non-zero radial velocity of the observer will contribute towards a dipole moment in CMB:
\be
{\de T\over T}\sim v_O=\dot{d_O}
\, ,
\ee
where the proper radial distance, $d_O$, of the observer  is given by
$$
d_O=\int_0^{r_O}dr\ {R'\over\sqrt{1+2(\bM r)^2k(r)}}
$$
Now, in our profile $k(r)$ remains almost a constant for almost the entire underdense region. Assuming we are living in this ``constant'' underdense region, we have
$$
d_O={2\pi(\cosh u-1)\over 3k_{\mt{max}}}\int_0^{r_O}{dr\over\sqrt{1+2(\bM r)^2k_{\mt{max}}}}={2\pi(\cosh u-1)\sinh^{-1}(\bM\sqrt{2k_{\mt{max}}}r_O)\over 3k_{\mt{max}}\bM\sqrt{2k_{\mt{max}}}}
$$
(The simplification occurs because $u$ and hence $R'$ becomes only a function of time.) Further, since $\bM r_O$ is expected to be very small, we have
\be
d_O={2\pi(\cosh u-1)r_O\over 3k_{\mt{max}}}
\, . \ee
Taking the time derivative and simplifying we find
\be
\dot{d_O}={d_OH_{\mt{out}}\over4}{u_0^3\sinh u\over(\cosh u-1)^2}
\label{ddot}
\, . \ee

We now note that $u(u_0)$ is a known function Eq.(\ref{f-defn}), in turn $u_0$  is known in terms of $\de_0$ via the function $f(u_0^2)$, see Eq.(\ref{f-de}). Thus, in principle, the second term in the right hand side of Eq.(\ref{ddot}) is determined in terms of the central underdensity contrast. Also, since the measured value of the CMB dipole moment $\sim 10^{-3}$, naturalness arguments suggest $ \dot{d_O}$ to be of the same order, and thus we have (after some simplifications):
\be
d_OH_{\mt{out}}\sim 10^{-3}{\sqrt{2}(1+f_0)^2\over\sqrt{u_0^2(1+f_0)^2+2(1+f_0)}}
\label{dipole-cons}
\, . \ee
For voids of around $200/h$ Mpc, and central underdensity contrasts between 40\% and 50\%, the dipole constraint Eq.(\ref{dipole-cons}) typically imply that ``we'' have to be located within 10\% of the void radius.

%%%%%%%%%%%%%%%%%%%%%%%%%%%%%%%%%%%%%%%%%%%%%%%%%%%%%%%%%%%%%%%%%%%%%%

\subsection{Analytic expression for the $D_L-z$ curve} \label{analyticEqs}
In this subsection we wish to provide the reader a self-consistent summary of all the equations which are needed to plot the $D_L-z$ curve, in an analytic form. Following this, a fit of any experimental dataset can easily be performed.
Here is the set of equations, which give $D_L$ and $z$ as a function of the radial coordinate $r$ (therefore implicitly $D_L-z$).
First of all one needs to define the function $f(u_0^2)$, implicitly given by:
\begin{eqnarray}
f & \equiv & \frac{\sqrt[3]{2} (\cosh (u)-1)}{3^{2/3} (\sinh (u)-u)^{2/3}}-1  \, \\
u_0 &=& 6^{1/3} (\sinh (u)-u)^{1/3}  \, .
\end{eqnarray}
Then, one can use this function in the following equations:
\begin{eqnarray}
\tau(r) &=& \tau_0-\frac{\pi}{9} \gamma^2 \bar{M} r [1+f(\gamma^2 \tau_0^2 k(r))] \, ,\\
1+z(r) &=& \left(\frac{\tau_0}{\tau(r)}\right)^2 \exp\left[ \frac{4 \pi \gamma^2 \bar{M} r}{9}f(\gamma^2 \tau_0^2 k(r)) \right] \, \\
D_L(r) &=& \frac{\pi}{3} \gamma^2 r \tau(r)^2 [1+f(\gamma^2 \tau_0^2 k(r))] [1+z(r)]^2 \, \\
\tau_0 &=& \left( \frac{2 \bar{M}}{3 H_{\rm out}} \right)^{1/3} \, \\
\gamma &=& \left( \frac{9 \sqrt{2}}{\pi} \right )^{1/3}
\end{eqnarray}

\begin{figure}
\includegraphics[width=0.48\textwidth]{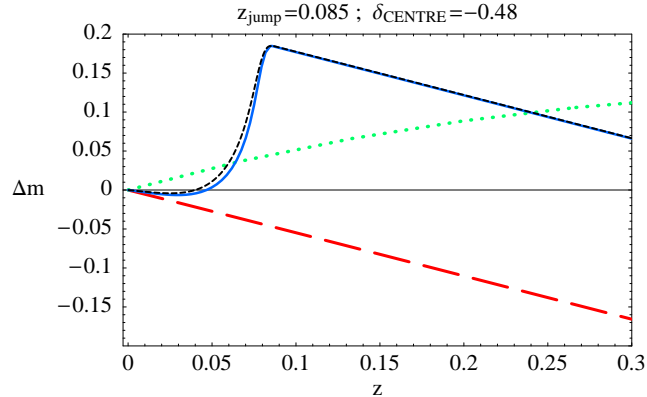}
\caption{ \label{fig:numanalytic} Comparison between analytic and numerical $D_L-z$ curves.
The numerical curve is the blue solid line, the analytic approximation is the black short-dashed line. We have plotted also the EdS curve (red long-dashed line) and the $\Lambda$ CDM, with $\Omega_{\Lambda}=0.7$ (green dotted line).
We have used the value $L=400$, with the units given in Eq.~(\ref{units}), and $k_{\mt{max}}=2.2$ (which corresponds to a density contrast at the center $\delta_0=-0.25$).}
\end{figure}

The above formulas are completely general for any LTB profile, but we now focus into our specific one given by
\be
k(r)=k_{\mt{max}}\LT1-\LF{r\over L}\RF^4\RT^2
\, . \ee
Then one has to choose appropriate values for $H_0$, and for the length units for the coordinate $r$ (given by $\bar{M}$).
A simple choice is to set:
\be
\sqrt{8\pi\over 3}\bar{M}=H_{\rm out}=h_{\rm out}/3000  \label{units}
\, ,
\ee
where we have chosen, in this way, the units Mpc=1 (which turns out to be a convenient choice for the problem).
Once this is done the physical parameter $L$ (the radius of the patch) is approximately given already in Mpc.
The comparison between the obtained curve and the fully numerical curve is shown in fig.~(\ref{fig:numanalytic})

%%%%%%%%%%%%%%%%%%%%%%%%%%%%%%%%%%%%%%%%%%%%%%%%%%%%%%%%%%%%%%%%
%%%%%%%%%%%%%%%%%%%%%%%%%%%%%%%%%%%%%%%%%%%%%%%%%%%%%%%%%%%%%%%%

Finally the reader may play with the two parameters: the size $L$ and $k_{\mt{max}}$ (which sets the amplitude of the density contrast).
We also recall that the density profile is given by Eq.(\ref{matterdensity}) and that $k_{\mt{max}}$ can be directly related to the density contrast $\delta_0$ at the center of the void at the present time, via the following equation:
\be
\delta_0=[1+f(\gamma^2 \tau_0^2 k_{\mt{max}})]^{-3}-1  \, .
\ee

%%%%%%%%%%%%%%%%%%%%%%%%%%%%%%%%%%%%%%%%%%%%%%%%%%%%%%%%%

\end{document}